\documentclass[twocolumn]{aastex631}

\newcommand{\kms}{\ifmmode{\,\hbox{km\,s}^{-1}}\else {\rm\,km\,s$^{-1}$}\fi}
\usepackage{verbatim}
\usepackage{hyperref}

\begin{document}

\shorttitle{M54 Dark Matter}
\shortauthors{Carlberg \& Grillmair}

\title{The Dark Matter Halo of M54}

\author[0000-0002-7667-0081]{Raymond G. Carlberg}
\affiliation{Department of Astronomy \& Astrophysics 
University of Toronto 
Toronto, ON M5S 3H4, Canada} 
\email{raymond.carlberg@utoronto.ca}

\author[0000-0003-4072-169X]{Carl J. Grillmair}
\affiliation{IPAC, California Institute of Technology, Pasadena, CA 91125}
\email{carl@ipac.caltech.edu}

\begin{abstract}
M54 is a prototype for a globular cluster embedded in a dark matter halo. {\it Gaia} EDR3 photometry and  proper motions separate the old, metal-poor stars from the more metal rich and younger dwarf galaxy stars. The metal poor stars dominate the inner 50 pc, with a velocity dispersion profile that declines to a minimum around 30 pc then rises back to nearly the central velocity dispersion, as expected for a globular cluster at the center of a CDM cosmology dark matter halo.   The Jeans equation mass analysis of the three separate stellar populations gives consistent masses that rise approximately linearly with radius to 1 kpc.  These data are compatible with an infalling CDM dark matter halo reduced to $\simeq 3\times 10^8 M_\sun$ at the 50 kpc apocenter 2.3 Gyr ago, with a central globular cluster surrounded  by the remnant of a dwarf galaxy. Tides gradually remove material beyond 1 kpc but have little effect on the stars and dark matter within 300 pc of the center.   M54 appears to be a \lq\lq{}transitional\rq\rq{} system between globular clusters with and  without local dark halos whose evolution within the galaxy depends on the time of accretion and  orbital pericenter.
\end{abstract}

\section{INTRODUCTION}

The relationship between the very old globular clusters  and sub-galactic mass cosmological dark matter halos \citep{DZN67,PeeblesDicke68,Peebles84}, is important to understand the earliest phases of star and galaxy formation. The observed internal kinematics of globular cluster stars over a range of stellar masses combined with point mass n-body modeling of individual clusters finds that the mass density of the clusters is consistent with the stellar mass of an old, evolved, stellar population inside 2-3 half-mass radii \citep{Zonoozi11,BaumgardtHilker18,Ebrahimi20,VB21}. At larger radii the declining density of cluster stars makes comtamination by foreground and background stars an increasing problem for velocity dispersion measurements. However, the growing number of radial velocity measurements and the increasing time baseline of the {\it Gaia} data \citep{GaiaMission2016}  is allowing the kinematics of  stars at larger radii to be examined.  

There is evidence that some, but certainly not all, halo globular clusters have velocity dispersion profiles at large cluster-centric radii above the predictions from the stellar mass of the cluster alone. Comparisons of the observed profiles to detailed n-body velocity dispersion profiles find that $\sim$10\% of currently measured clusters have velocity dispersion profiles  above the values predicted from the stars \citep{BaumgardtHilker18,BianchiniIF19,Wan_etal21}. For a set of 19 halo clusters measured to at least 5 half mass radii about 10\% have a rising velocity dispersion profile and another 30\% are approximately flat \citep{CarlGrill21}. The current measurements are still somewhat noisy but will improve significantly with future {\it Gaia} releases. 

There are a number of possible explainations for some or all of the excess velocity dispersion, including single epoch observations of binary stars \citep{Wan_etal21}, the complexity of orbits between the cluster and the tidal surface \citep{FukushigeHeggie00,ClaydonGZ17}, particularly for highly elliptical orbits near the galactic center \citep{CarlGrill21}, stripped galactic nuclei \citep{KuzmaDaCM18,Wirth20}, and dark matter halos \citep{MS05a,MS05b,Penarrubia17,Boldrini20,VitralBondrini21,CK22}. Dynamical measurements of potential progenitor systems can be used to test these ideas.

The  M54/NGC6715 cluster is at the center of the Sagittarius (Sgr) dwarf galaxy \citep{Ibata94,Bellazzini08}. M54, with a mass of $1.4-1.8\times 10^6 M_\sun$, is one of the most massive globular clusters within the Milky Way halo. M54 satisfies the definition of nuclear star cluster (NSC), one of four within smaller galaxies that have been identified as having accreted into the Milky Way \citep{Pfeffer21}. As a nuclear star cluster M54 is around the median NSC mass and one of the most metal poor known \citep{NSCReview2020}. The stars in M54 are predominantly old and metal poor \citep{Bellazzini08,AlfaroCuello19} as opposed to several nearby NSCs whose central regions are predominantly young stars \citep{Hannah21}. M54 has no compelling evidence for a central black hole \citep{Ibata09,Wrobel11}. The Sagittarius dwarf luminosity, $L_V=1.8\times 10^6 L_\sun$ \citep{Mateo98} is a factor of 10 less luminous than expected, relative to the mean relation for NSC hosts \citep{NSCReview2020}.  The Sgr dwarf has  prominent tidal tails \citep{Ibata97,Belokurov06:FieldofStreams,Ibata20:SgrStream} which help constrain the recent orbital history of the system.  Spectroscopic observations of the line of sight velocity profile of M54 find that the old, metal poor, blue stars of M54 have a velocity dispersion profile that declines from a central value of approximately 10 \kms\  to a minimum of about 5 \kms\ at 22 pc and then rises back up to nearly 10 \kms\ again at 55 pc, where it becomes essentially the same as the velocity dispersion of the younger, more metal rich, redder stars of the dwarf galaxy surrounding the cluster \citep{Bellazzini08,AlfaroCuello20}. 

This paper uses {\it Gaia} data to separate the stars within about 4 kpc of M54 into blue, red and an intermediate color (\lq\lq{}green\rq\rq{}) populations. The proper motions give the projected radial and tangential velocities of the stars. The resulting radial velocity dispersion profiles and the degree of radial anisotropy of each distinct stellar population are used  with the Jeans equation to measure the total mass profile with radius. The results are compared to n-body models to constrain the dark mass near and into the star cluster and within the surrounding dwarf galaxy. The models are also used to assess the long term evolution of the M54 star cluster-dark halo systems and its dependence on the system\rq{}s orbital pericenter. 

\section{Data}

We use the Early Data Release 3 \citep{GaiaEDR3} of the {\it Gaia} mission \citep{GaiaMission2016} to select $B_P - R_P$ colors, G magnitudes, parallaxes, proper motions, and their estimated errors, for stars within 10 degrees of the globular cluster M54/NGC6715, or 4.587 kpc  at our adopted distance of 26.28 kpc \citep{BV21}. The Gaia photometry is corrected for extinction using the reddening maps of \citep{SchlegelFinkDavis98}, themselves corrected using the prescription of \citet{SchlaflyFinkbeiner11}, and using the Gaia DR2 coefficients derived by \citet{GaiaDR2HRDiagrams}. A $G_0, (G_{BP} - G_{RP})_0$ color-magnitude locus for the red giant branch (RGB), subgiant branch, and main sequence is constructed using stars between one and three arcminutes from the cluster center. A theoretical isochrone with [Fe/H] = -1.59 \citep{Harris96}  generated in the Gaia passbands at http://stev.oapd.inaf.it/cgi-bin/cmd \citep{Girardi04} is used as a guide at fainter magnitudes. This isochrone  is not used directly due to significant and variable offsets between the observed and theoretical giant branches. Color offsets from this sampled color-magnitude locus are then computed using the tabulated photometric uncertainties in Gaia EDR3.  The proper motion uncertainties lead to undertainties in each velocity component of 5 and 10 \kms\  at $G_0$=15.5 and 17.0 mag, respectively. The data are limited to an extinction corrected brightness of  $G_0$=17.5 magnitudes. Hereafter we drop the 0 subscript for the extinction corrected magnitudes and colors.

\begin{figure}
\begin{center}
\includegraphics[scale=0.45,trim=10 30 0 20, clip=true]{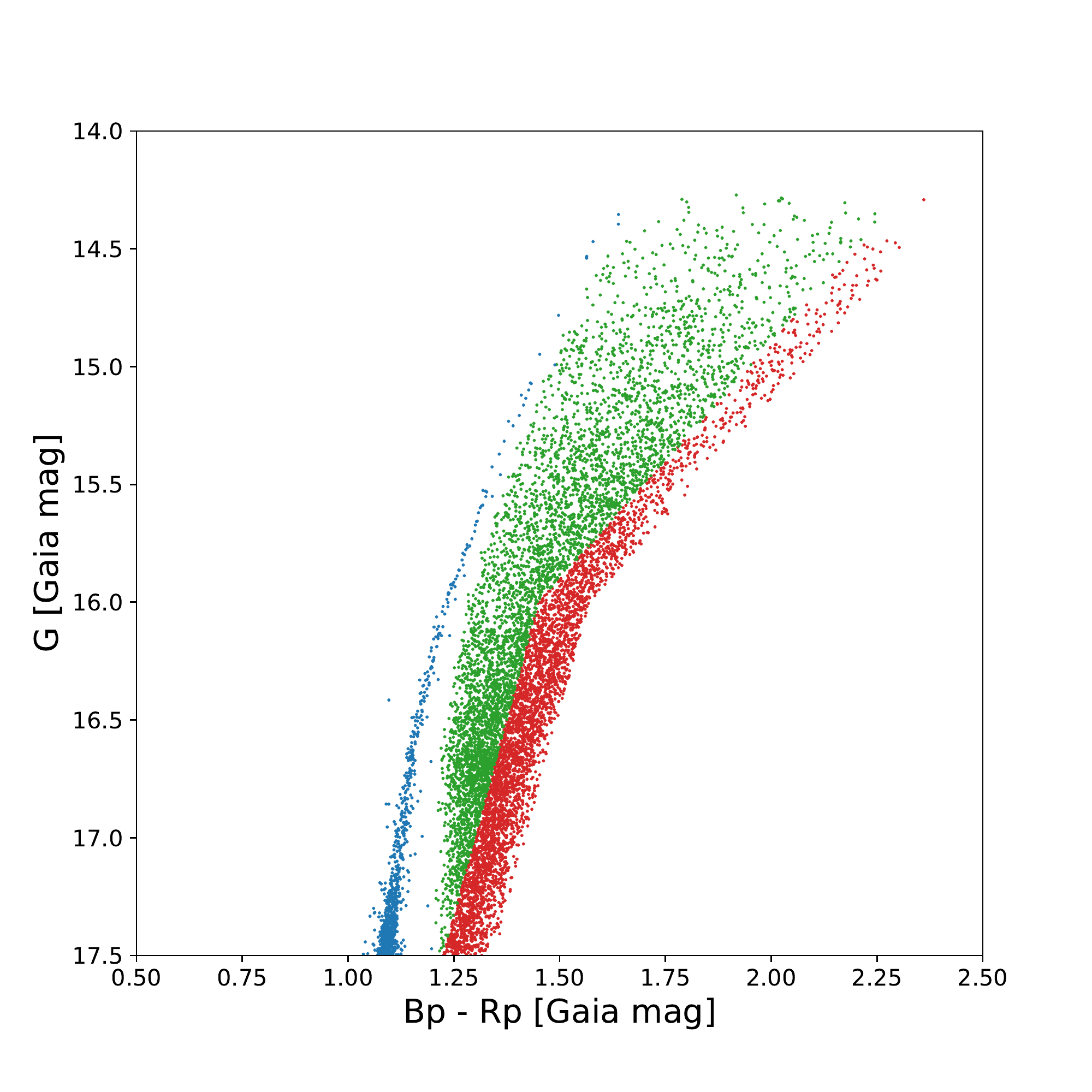}
\end{center}
\caption{The extinction corrected color-magnitude diagram for the velocity selected star sub-samples.}
\label{fig_cm}
\end{figure}

There are several distinct age and metallicity stellar populations within the M54 cluster and Sagittarius dwarf \citep{Bellazzini08,Ibata09,VB20:Sgr}. Detailed spectroscopic studies find that the cluster stars are predominantly an old metal poor population with a continuous spread to a younger metal rich population \citep{AlfaroCuello19}. Along the lines of previous studies \citep{Bellazzini08,delPino21} we use the Gaia photometry to separate the stars into three groups: the blue population, which is old and metal poor, a red population which is young and metal rich, and a population with intermediate colors that we designate \lq\lq{}green\rq\rq{}. The blue stars are those having a color within 1 sigma of the cluster sequence. The red stars are defined as those between $ B_P-R_P>1.45 + 0.15 (16-G)$ and $B_P-R_P<1.57+0.15(16-G)$ for $18>G>16$, and for $G<16$ colors between $B_P-R_P>1.45 + 0.5 (16-G)$ and $B_P-R_P<1.57+0.5(16-G)$. The intermediate color, \lq\lq{}green\rq\rq{}, stars are those that are at least 10 sigma to the red of the cluster sequence and bluer than the blue edge of the red sequence. The resulting color-magnitude diagram is shown in Figure~\ref{fig_cm}. The sample shown in Figure~\ref{fig_cm} includes cuts on velocity relative to the cluster center of mass, as discussed below.

\begin{figure}
\begin{center}
\includegraphics[scale=0.45,trim=20  0 0 0, clip=true]{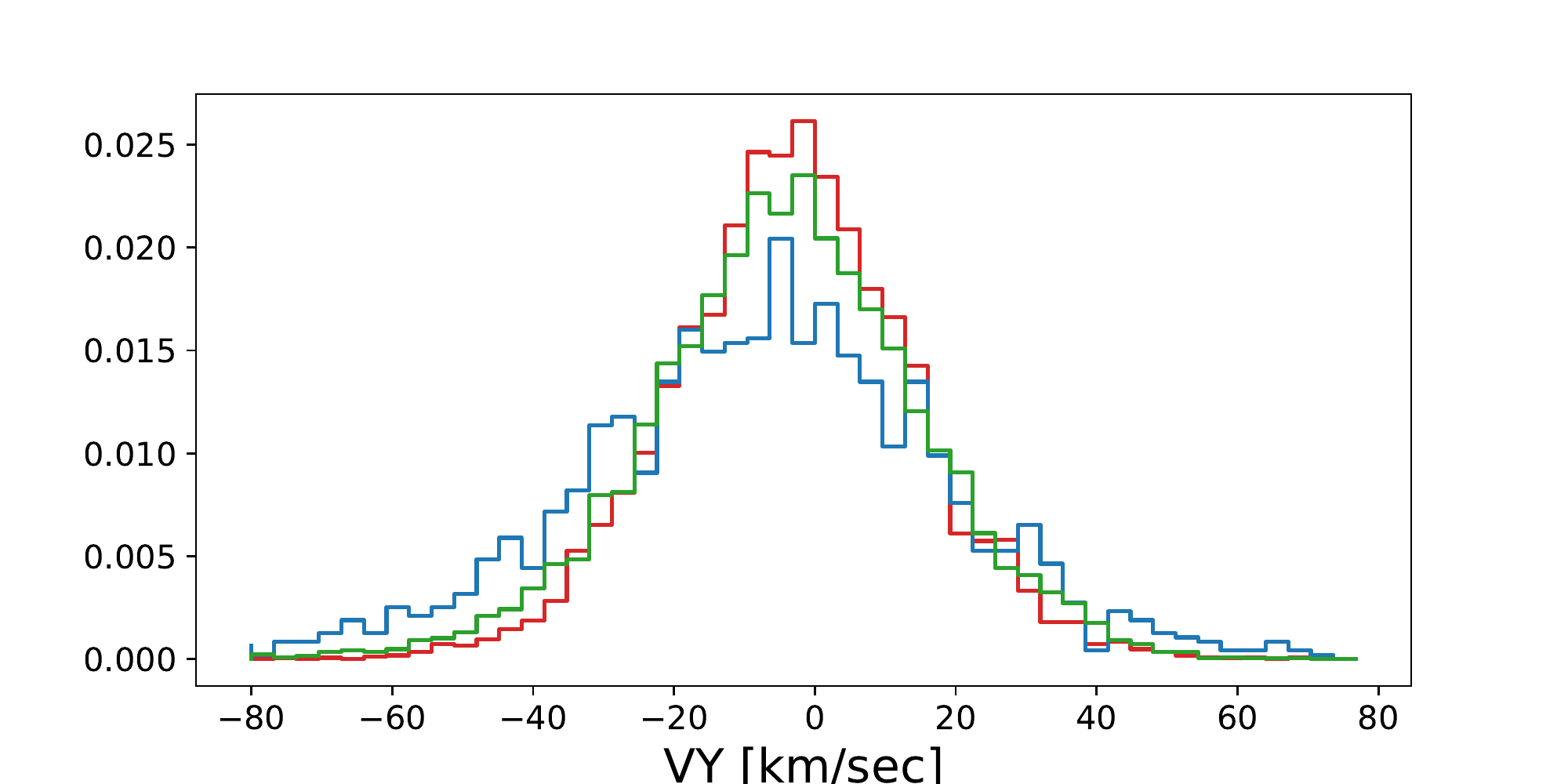}

\includegraphics[scale=0.45,trim=20 0 0 32, clip=true]{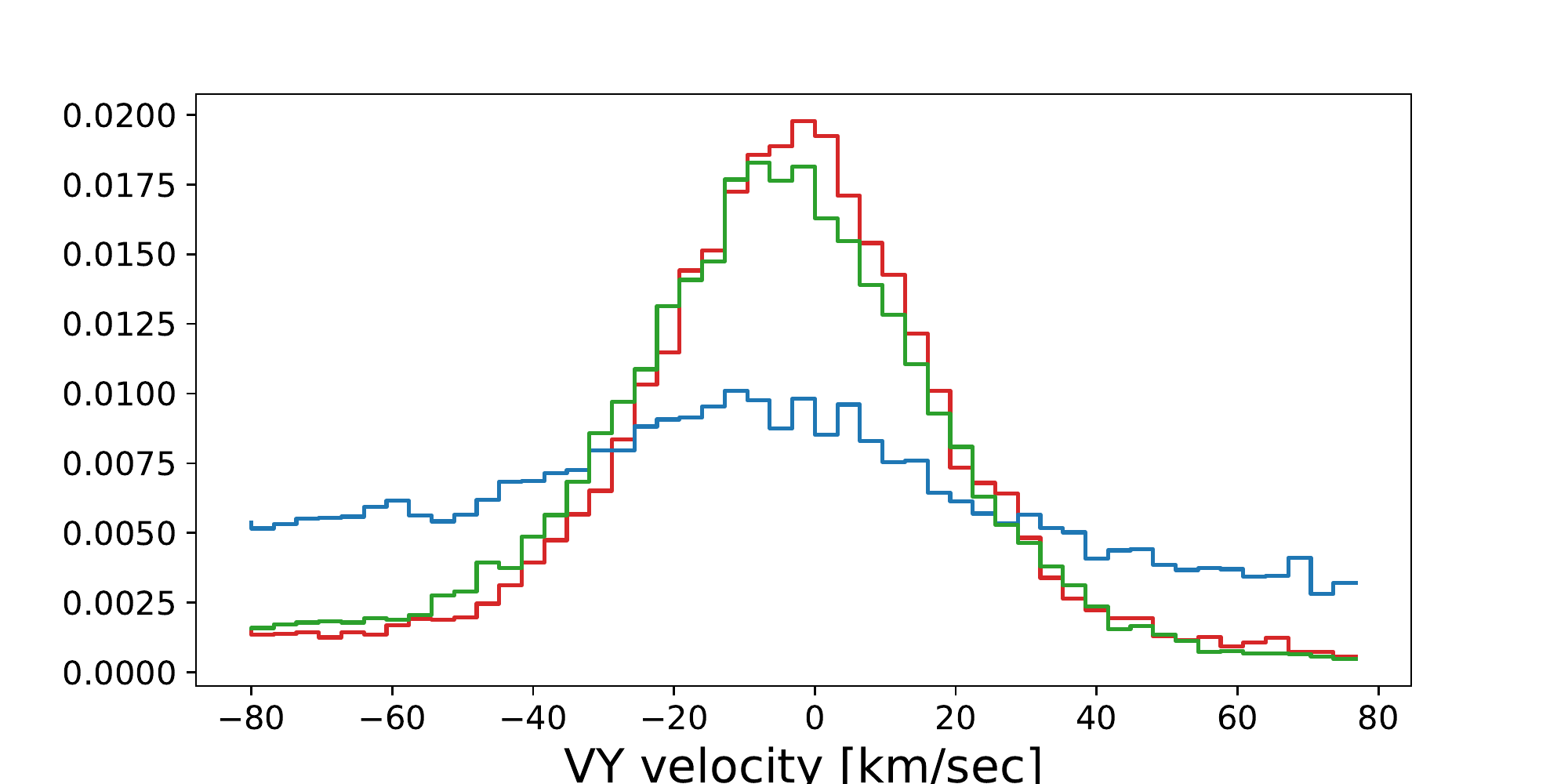}
\end{center}
\caption{The unit area normalized distribution of relative stellar velocities parallel to the galactic plane for the three color sub-samples.  The top panel is for stars within 1 kpc of the cluster, the bottom for 4 kpc. The velocities perpendicular to the plane have similar distributions. }
\label{fig_vdist}
\end{figure}

The M54 cluster is located close to the Galactic plane, at Galactic latitude of -14.1\degr\ and longitude 5.6\degr\, resulting in a large and variable foreground of Galactic stars.  The sample also has systematic errors with position \citep{delPino21}.  The parallax errors are sufficiently large for most of the stars that a parallax limit is not helpful.  Velocities are calculated assuming that the stars are at the distance of the cluster and placed in the center of mass frame of the cluster. Velocity cuts are important to reduce both the background and systematic errors.

\begin{figure}
\begin{center}
\includegraphics[scale=0.5,trim=10 35 0 20, clip=true]{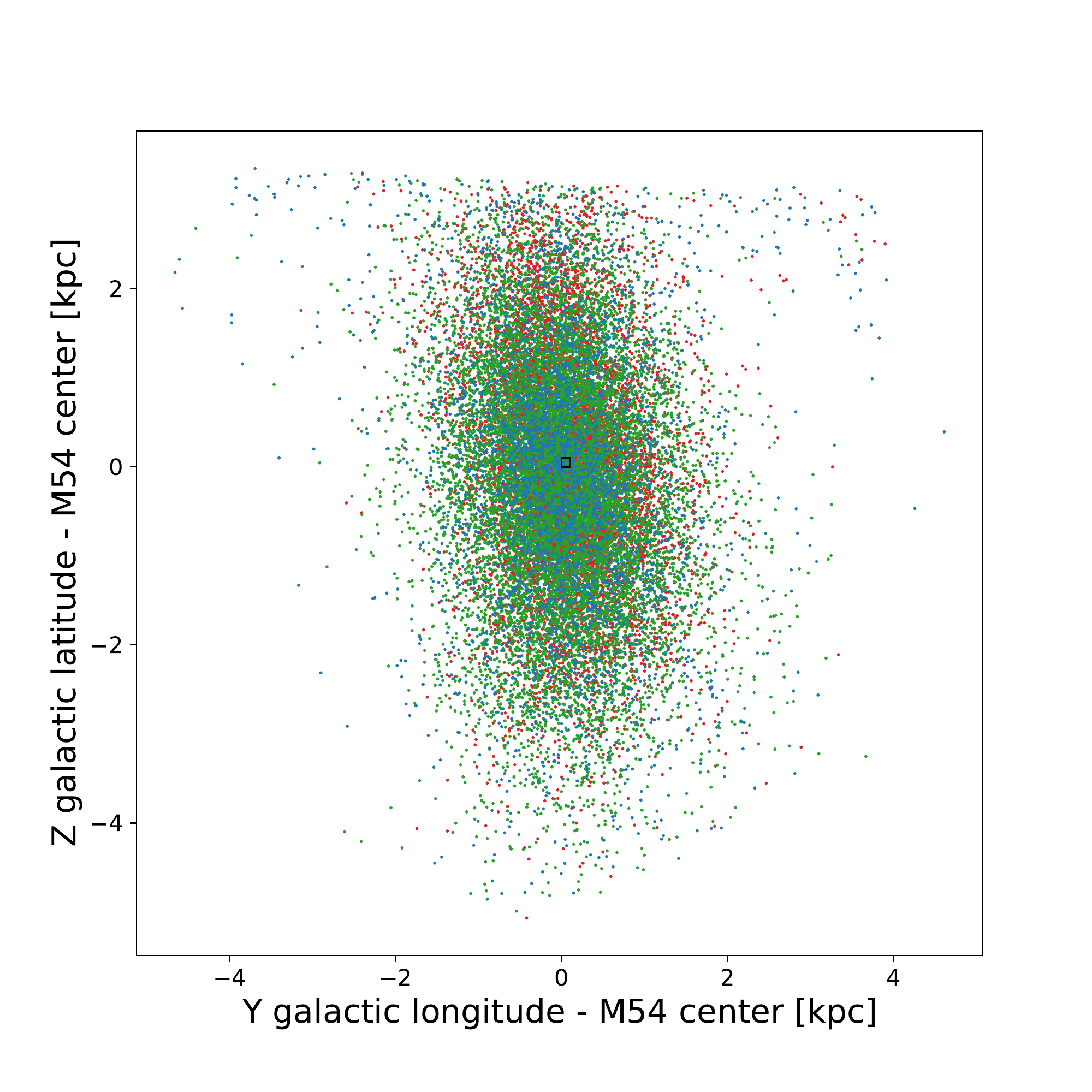}
\put(-245,18){\includegraphics[scale=0.15,trim=0 0 0 0, clip=true]{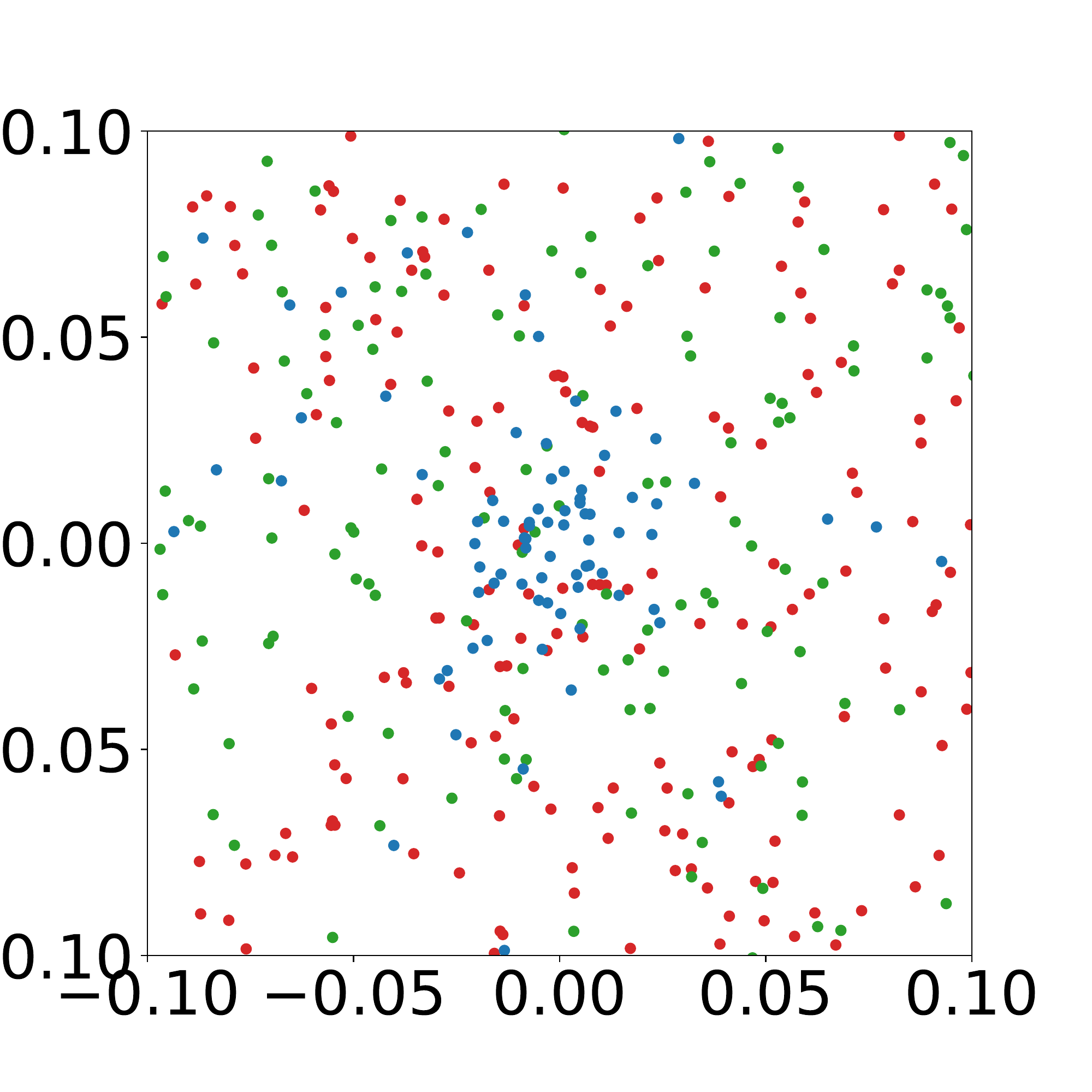}}
\end{center}
\caption{The sky distribution of the  M54/Sgr $|v_{XY}|\leq 30 \kms$ stars plotted in our longitude and latitude coordinates converted to kiloparsecs at the adopted distance of M54. The Galactic mid-plane is at y=6.46 kpc in these coordinates. The inner $\pm$0.1 kpc region, marked with a square at the center, is shown in the inset.  At the distance of M54, 1 kpc is 2.18\degr\ or 7.64 pc per arc-minute. Crowding causes an increasingly large under-sampling in the central region, where the M54 stars dominate those of the Sgr dwarf.}
\label{fig_M54xy}
\end{figure}

The radial velocities in the plane of the sky are projection corrected, using the center of mass velocity of M54, 143.06 \kms\ \citep{Baumgardtetal19} at a consensus distance of 26.28 kpc  \citep{BV21}. We will designate the two coordinates in the plane of the sky Y and Z, which are close but not quite identical to the Galactic Y and Z. Relative to the Galactic frame, our Z coordinate is  tilted 5.5\degr\ backward from us and our  Y coordinate is rotated 14\degr\ counterclockwise, as seen from above. The distribution of the velocities  parallel to the Galactic plane, relative to the cluster, is shown in Figure~\ref{fig_vdist} for stars within 1 and 4 kpc of the cluster (top and bottom panels, respectively). There is a clear excess of stars above 30 \kms\ at large radii.  The velocities perpendicular to the plane show a similar excess of blue stars with velocities  larger than about 30 \kms\ beyond 1 kpc, most of the stars being those close to the plane. Gaia data has various sky location selection effects in the region of M54 (see \citet{delPino21} for a discussion of the DR2 data). To produce an acceptably uniform sample in the M54 region we  select stars that have a projected velocity relative to the cluster of 30 \kms\ or less. The distribution of the selected stars on the sky is shown in Figure~\ref{fig_M54xy}.  While Galactic plane stars intrude at the top of the figure,  background contamination has been greatly reduced. Using a larger velocity cut of 50 \kms\ makes little difference for densities but  boosts velocity dispersions at all radii, with a strong rise beyond 300 pc. We adopt the 30 \kms\ limit for our measurements.

\section{M54-Sgr Density and Velocity Dispersion Profiles}

\begin{figure}
\begin{center}
\includegraphics[scale=0.45,trim=10 30 0 0, clip=true]{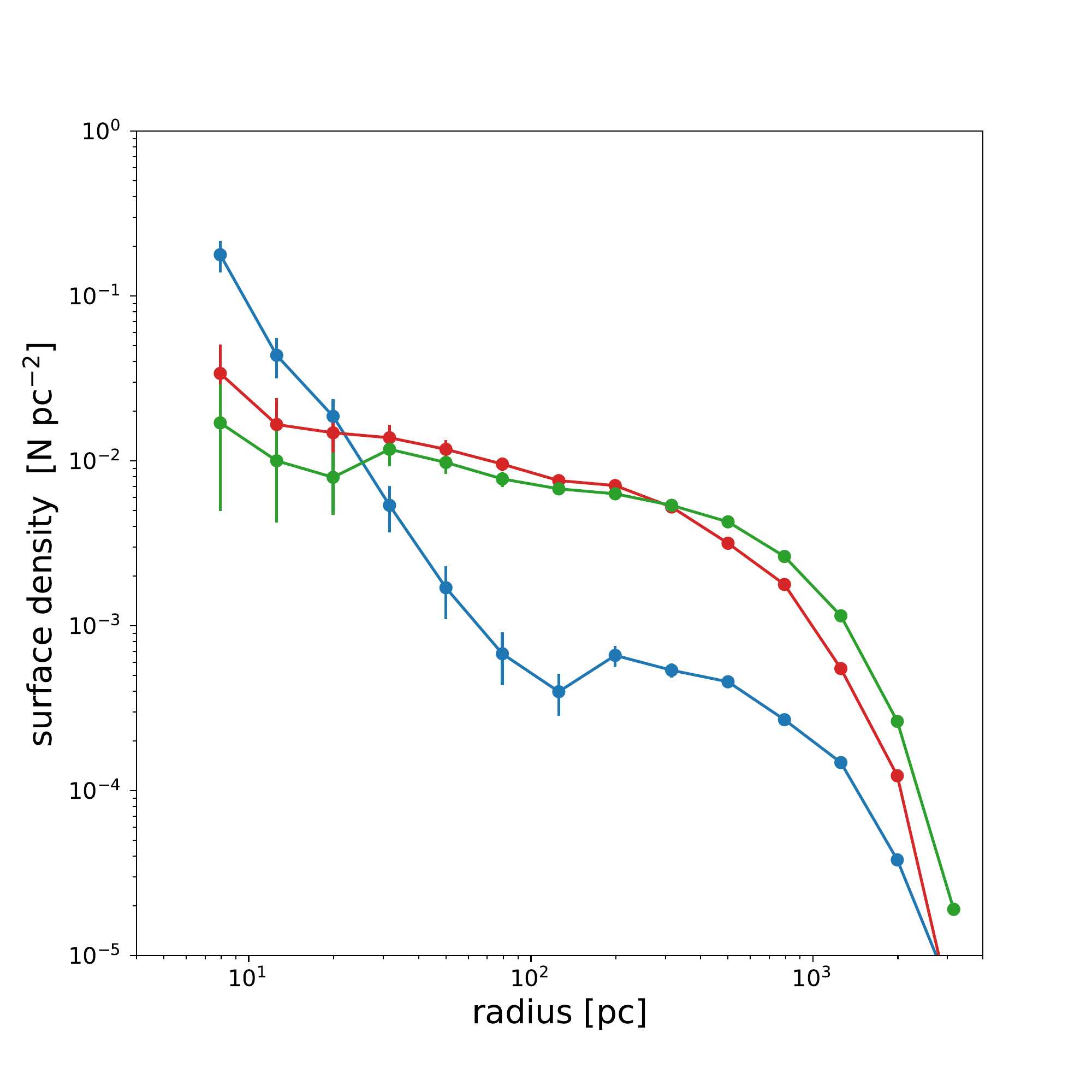}
\end{center}
\caption{The azimuthally averaged surface density versus projected radius for the color selected stars with absolute velocities less than 30 \kms. Image crowding greatly reduces completeness inside approximately 20 pc. The central M54 cluster has a 3D half mass radius of 5.6 pc. }
\label{fig_denr}
\end{figure}

The density and velocity dispersion profiles of  the star cluster are measured as projected azimuthally averaged quantities, which is appropriate for the M54/NGC6715 as it has an ellipticity of 0.06 \citep{Harris96}. Beyond 100 pc the dwarf galaxy stars are in a significantly elliptical distribution, as shown in Figure~\ref{fig_M54xy}. The Sagittarius dwarf galaxy is  modeled as a prolate triaxial dwarf galaxy with its axis of rotation tilted approximately 40-60\degr\ from the sky plane \citep{delPino21} with gradients of both density and velocity dispersion  so shallow that radial averaging is acceptable.  The radially averaged surface densities of the three color samples are shown in Figure~\ref{fig_denr}.

The red and green stars dominate the  surface density beyond 50 pc in the Sgr dwarf, with  blue stars comprising about 5\% of the combined surface density beyond 100 pc.  The central part of the field is so crowded that the density profile of M54 is severely undersampled inside 10 pc \citep{Pancino17}, nearly twice M54\rq{}s half-mass radius of 5.6 pc. The green and red subsamples show only a weak rise towards the center in these data. Better resolved data for the young and metal rich red  stars finds that their density rises within the cluster approximately in proportion to the blue stars, but with about 10\% of the surface density \citep{Bellazzini08}.  Sgr is unusual in having a dominant old, metal poor population, whereas young, metal rich stars often dominate the central regions of well  resolved nearby nuclear star clusters \citep{Hannah21}. 

\begin{figure}
\begin{center}
\includegraphics[scale=0.48,trim=20 30 0 0, clip=true]{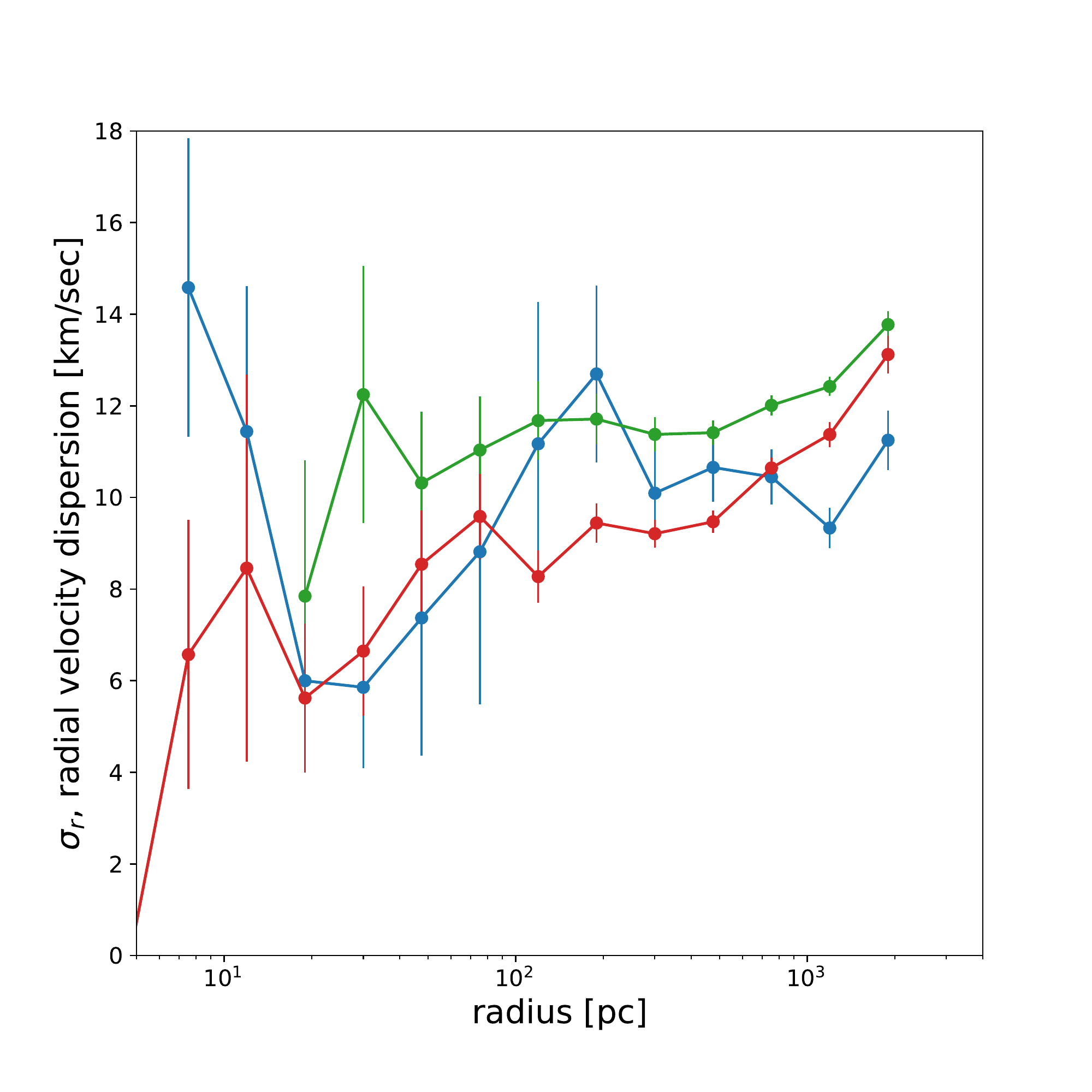}
\put(-132,23){\includegraphics[scale=0.25]{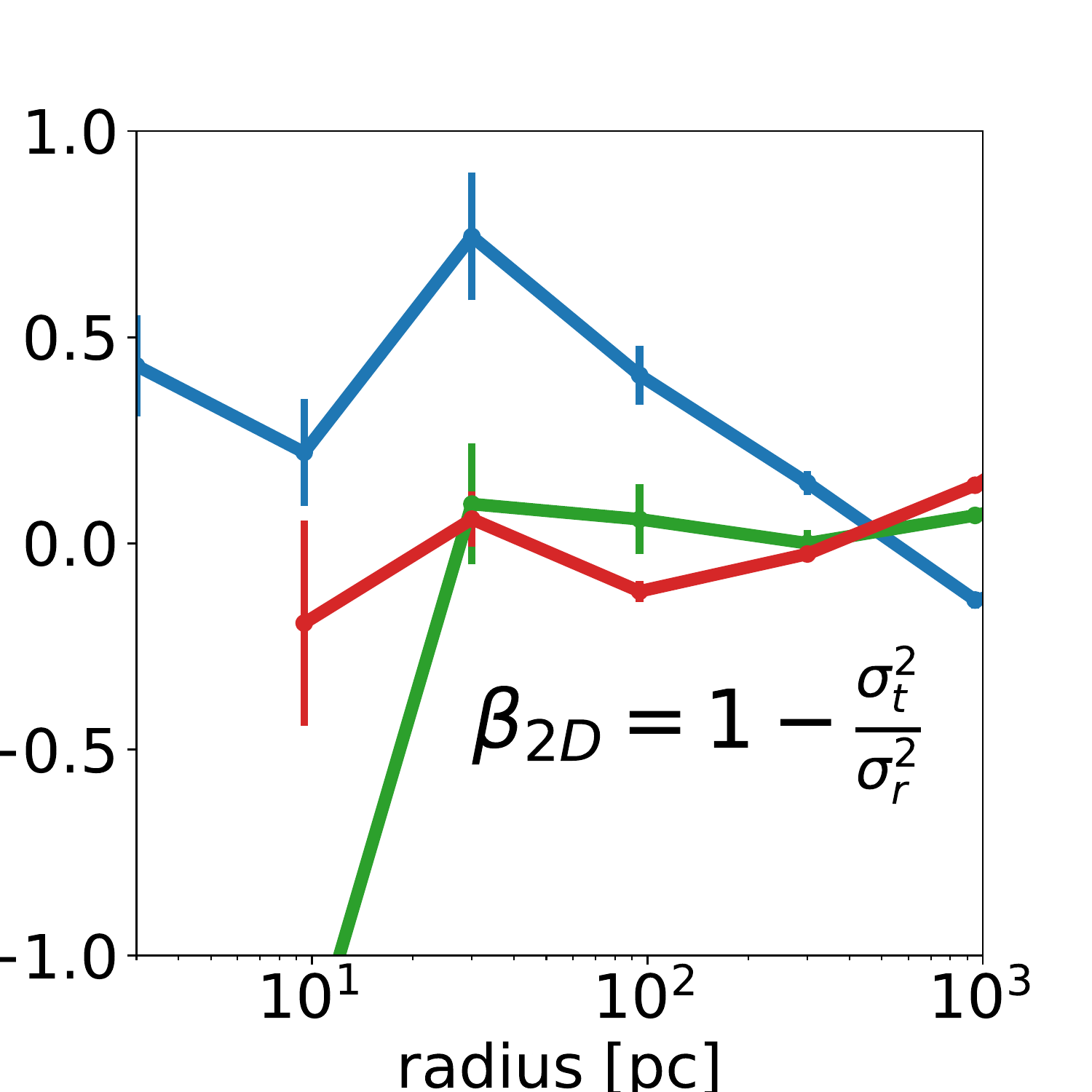}} 

\includegraphics[scale=0.48,trim=20 30 0 0, clip=true]{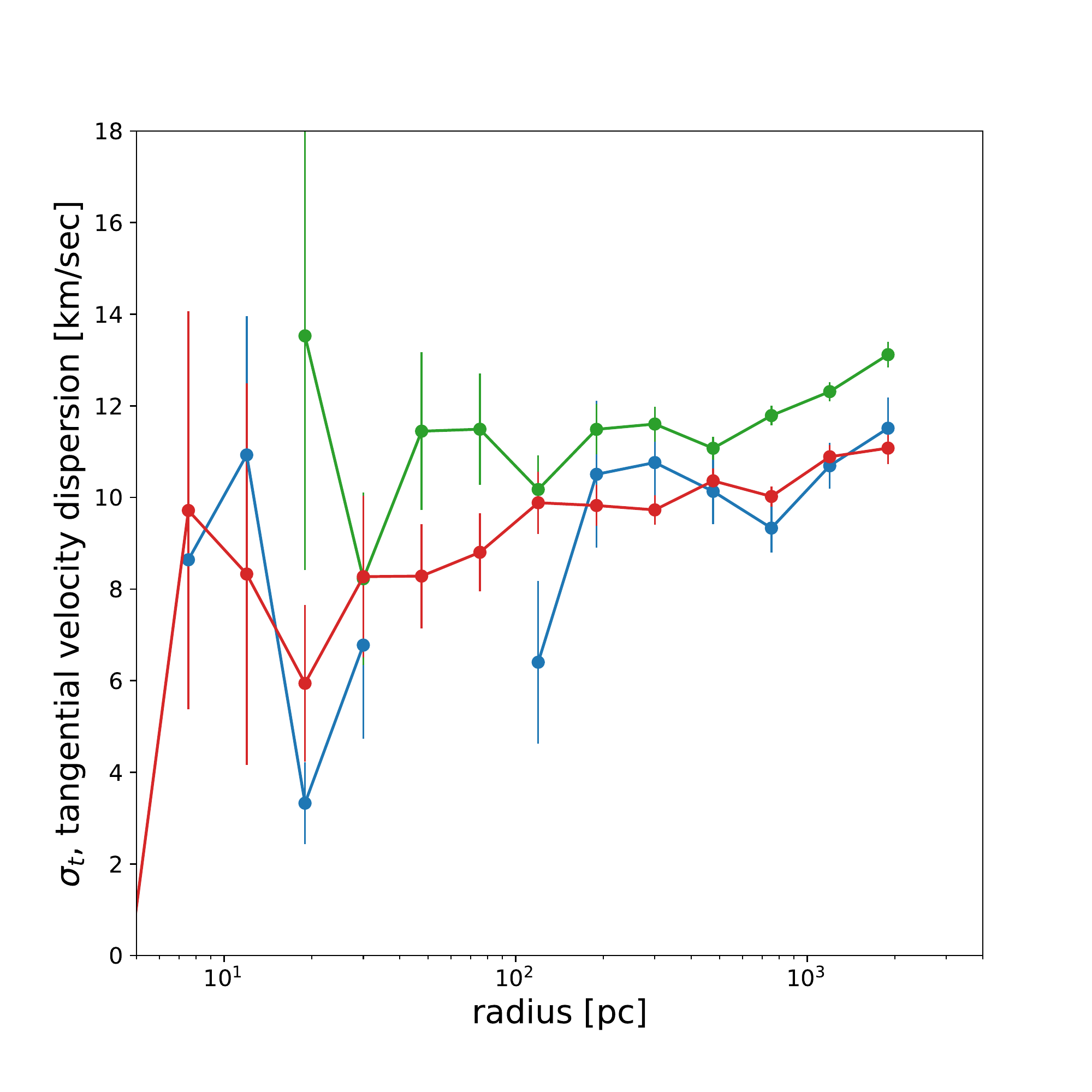}
\end{center}
\caption{The error corrected projected radial (top panel) and tangential (bottom panel) velocity dispersion profile of M54/Sgr for the blue, green, and red subsamples. The 2D velocity anisotropy,  $\beta_{2D} = 1- \sigma_t^2/\sigma_r^2$, evaluated in wider radial bins, is shown in the inset using the same line colors. The tangential velocity dispersion between 50 and 100 pc, where there are less than 10 data points per smaller bin, is consistent with zero}
\label{fig_M54sigr}
\end{figure}

The 2D velocity dispersion in the plane of the sky is $\sigma_{2D} = \sqrt{\sigma_r^2+\sigma_t^2}$, where $\sigma_r$ and $\sigma_t$ are the velocity dispersion measured in the plane of the sky relative to the cluster center.  The velocity dispersions have the  mean proper motion velocity error in each bin,  which range from about 8 to 12 \kms, subtracted  in quadrature.  The resulting projected radial  velocity dispersion, $\sigma_r(r)$ and tangential velocity dispersion,  $\sigma_t$, of the three color subsamples are shown in Figure~\ref{fig_M54sigr}. The tangential velocity dispersion in the range of 50 to 100 pc is consistent with zero, with less than 10 data points per bin in this radial range making the measurement of a small velocity dispersion difficult. The combined 2D velocity dispersion is shown in Figure~\ref{fig_M54sigstar} compared with the n-body velocity dispersions.

The red and green stars, primarily in the dwarf galaxy, have similar velocity dispersion profiles, with a shallow decline toward the center inside 1000 parsecs. The blue stars, primarily in the central star cluster, have a distinctive velocity profile, with a minimum at 30 parsecs and beyond that a rise back up to nearly the central value.  The velocity dispersion dip in the blue stars and the flat profile of the red stars was noted in the line-of-sight radial velocity data of \citet{Bellazzini08};  the data here confirm that behavior using the orthogonal components provided by the Gaia proper motions.  The lower velocity dispersion of the red stars relative to the intermediate green stars was seen in the line-of-sight velocity analysis of \citet{AlfaroCuello20}.

The inset in Figure~\ref{fig_M54sigr} shows the 2D velocity anisotropy, $\beta_{2D} = 1- \sigma_t^2/\sigma_r^2$, in somewhat wider radial bins to increase the signal-to-noise ratio. The velocity ellipsoid of the blue stars becomes nearly completely radial around 80 pc. At larger radii all three color subsamples have nearly isotropic velocity  ellipsoids. The blue stars in the central 50 pc  are kinematically distinct from stars of all colors at larger radii, which belong to the Sagittarius dwarf galaxy and have ages ranging from the last 0.5 Gyr to a Hubble age \citep{AlfaroCuello19}. 

\section{Total Mass Profile within 1 kpc}

\begin{figure}
\begin{center}
\includegraphics[scale=0.45,trim=10 30 40 45, clip=true]{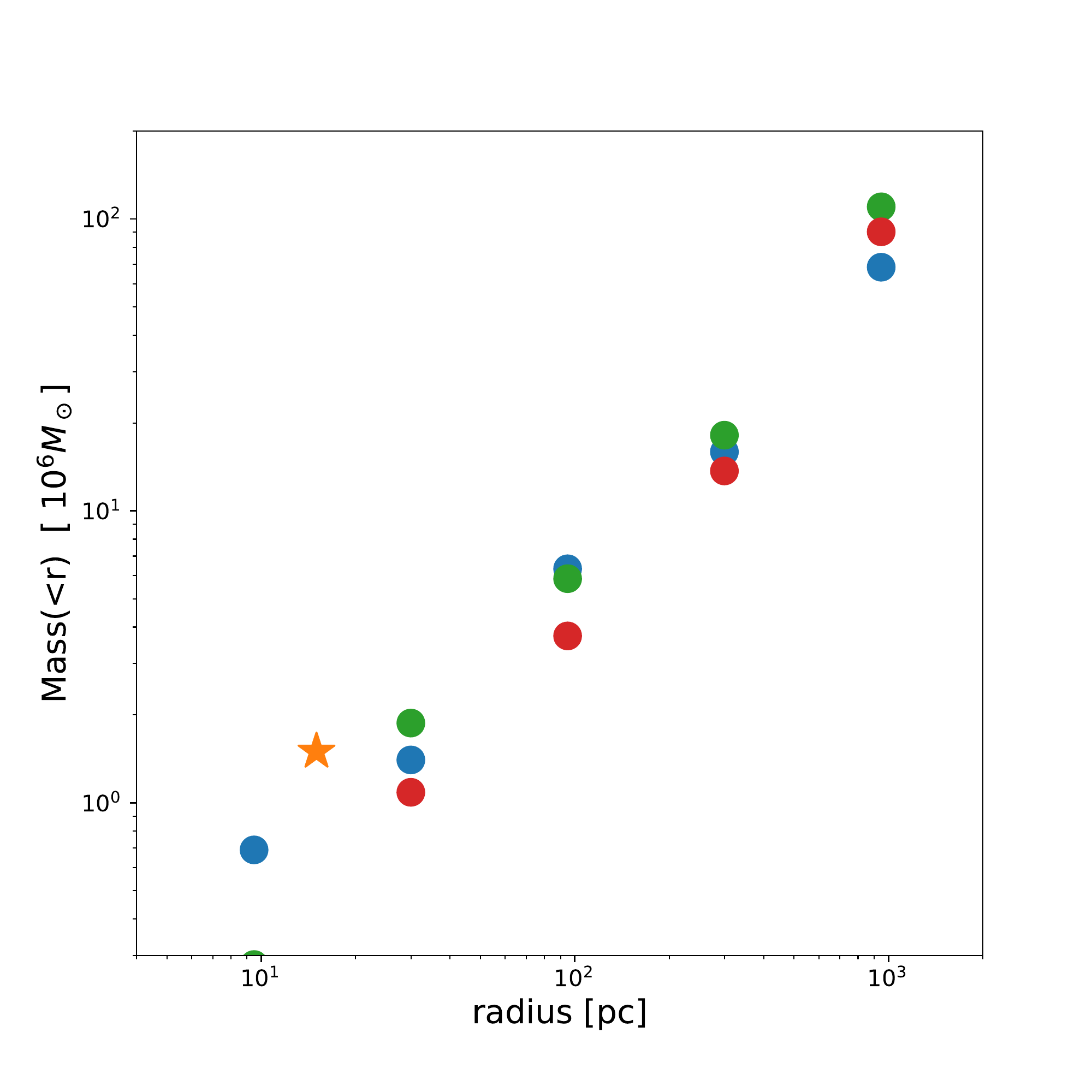}
\end{center}
\caption{Mass estimates from the Jeans equation for the three independent color subsamples. The mass points partially overlap at the largest radii. The star symbol indicates the \citet{BaumgardtHilker18}  stellar mass of the cluster which is placed at 3 half-mass radii.}
\label{fig_jeansmass}
\end{figure}

The radial Jeans equation measures the total mass interior to radius $r$,
\begin{equation}
M(r) = - {{r \sigma_r^2}\over{G}} \left[ \frac{d \ln{\nu}}{d\ln {r}} + \frac{d \ln{\sigma_r^2}}{d\ln {r}} + 2 \beta(r)   \right],
\label{eq_Jeans}
\end{equation}
where  $\nu(r)$ is the density of a tracer population, $\sigma_r(r)$ is its radial velocity dispersion, and $\beta(r)$ is its 3D velocity dispersion anisotropy. The data displayed in Figures~\ref{fig_denr} and \ref{fig_M54sigr} are used in the Jeans equation to estimate total interior masses shown in Figure~\ref{fig_jeansmass}.  The Jeans equation assumes that the stars are in equilibrium and we are making the assumption of a spherical underlying potential. These assumptions are examined with models below.

The power-law slope, $s_\nu=d \ln{\nu}/d\ln {r}$, of the 3D stellar density profiles, $\nu(r)$, is estimated from the slope of the surface density profile of the population, using the power law relation that the 3D slope is the 2D slope plus one.  The 2D $\beta$ values from the Figure~\ref{fig_M54sigr} inset are  close to the 3D values,  $\beta_{3D} = 1- (\sigma_{t1}^2+\sigma_{t2}^2)/2\sigma_r^2$, for our assumed spherically symmetric system in which the two components, $\sigma_{t1}$ and $\sigma_{t2}$, of the 3D tangential velocity dispersion, are equal. For the two special cases of the velocity ellipsoid being isotropic or completely radial, $\beta_{2D}=\beta_{3D}$, with values of 0 and 1, respectively. 

Figure~\ref{fig_jeansmass} shows the results of the Jeans mass analysis. An important outcome is that the three separate star samples give consistent results at 30 pc and beyond. The mass inside 30 pc is estimated to be  1.4, 1.9 and 1.1$\times 10^6 M_\sun$ for blue, green and red stars, respectively. The 30 pc mass is consistent with the stellar dynamical mass of the star cluster alone, though it does not rule out a small fraction of dark matter.  Between 30 and 95 pc the total interior mass rises from the average value of $1.5\times 10^6$ to $3.5\times 10^6 M_\sun$, a factor 2.3. There is little increase in the stellar mass in this region, so the bulk of this mass increase must be made up of dark matter. The mass rises another factor of about 17 from 95 to 950 pc to $89\times 10^6 M_\sun$. Again, this must be made up mostly of dark matter,  though the stars of the Sagittarius dwarf (with about 70\% of the  luminosity $L_V=1.8\times 10^7 L_\sun$ within 950 pc) will contribute  about 15\% of the mass for a stellar $M/L_V\simeq 1$.

\section{Comparison with N-body models}

\begin{figure}
\begin{center}
\includegraphics[scale=0.45,trim=10 30 40 45, clip=true]{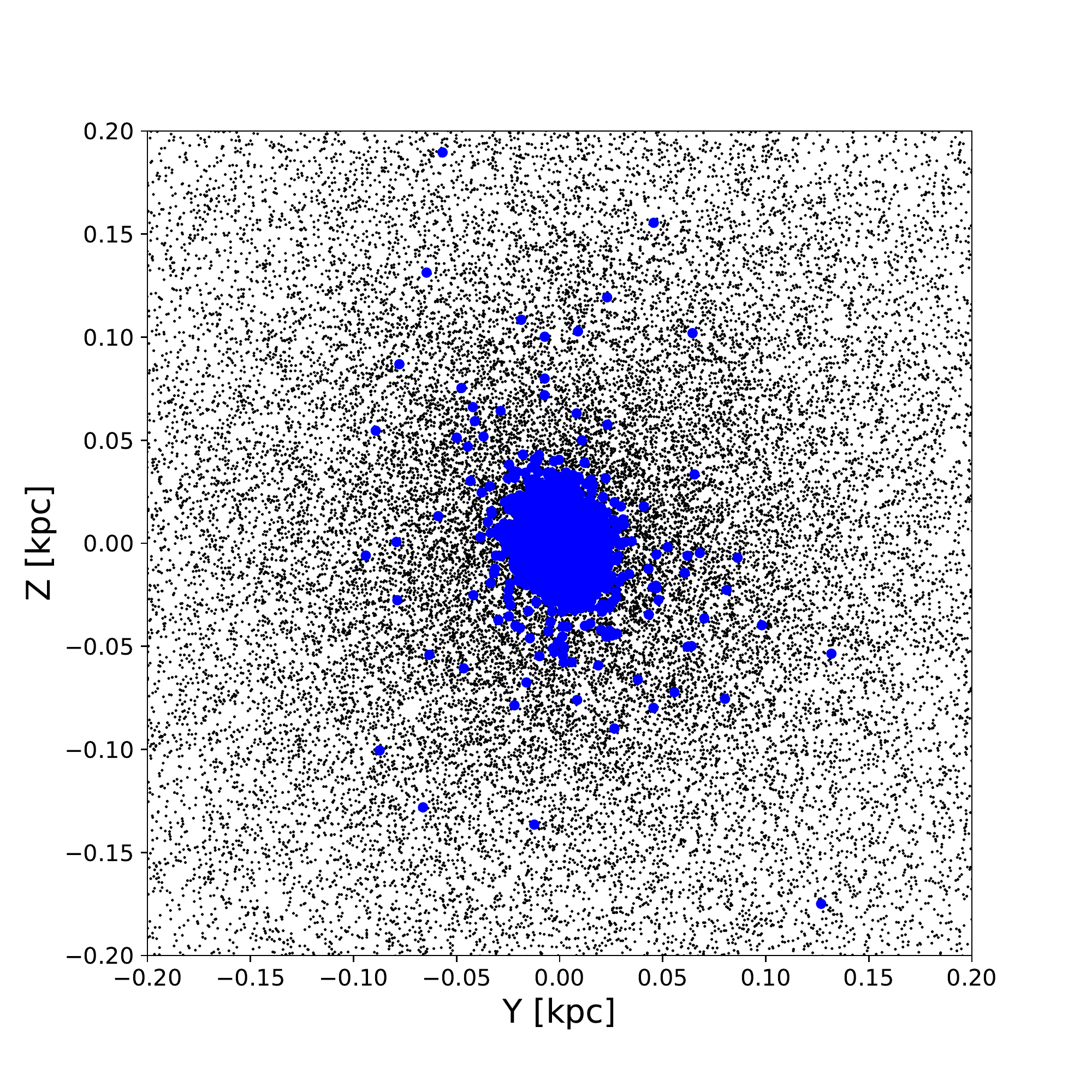}
\end{center}
\caption{The n-body model cluster stars (blue) and one of every 50 dark matter particles in the $3\times 10^8 M_\sun$ halo simulation.  }
\label{fig_YZstar}
\end{figure}

Our model for the M54-Sgr system begins when the system has settled into its current orbit, likely having shed the outer part of a more massive dark matter halo \citep{Lokas10}. A King model star cluster \citep{King66} is inserted into a Hernquist model dark halo \citep{Hernquist90}. The two populations equilibrate in 0.02 Gyr. The combined star cluster-dark halo system is placed at the apocenter of M54\rq{}s orbit and integrated backward in time in a MW2014 potential \citep{Bovy15} with the bulge component modeled as a 0.5 kpc Plummer sphere. The large mass-to-light ratio of the Sagittarius dwarf, 22 $M_\sun/L_\sun$ \citep{Mateo98} or the 15\% within 1 kpc we found in our Jeans mass analysis means that the  stellar mass of the dwarf galaxy contributes little to the overall potential and can be ignored.  The initial system is spherically symmetric. We  select an inner, flattened, tilted, subset of dark matter particles on relatively circular orbits as proxies for the dwarf galaxy disk stars to follow their evolution with time. 

\subsection{The N-body Model Star Cluster}

We adopt a star cluster mass of $1.78\times 10^6 M_\sun$ and a  half-mass radius of 5.6 pc for M54 \citep{BaumgardtHilker18} (https://people.smp.uq.edu.au/ HolgerBaumgardt/globular).  The star cluster along with the surrounding dark matter is integrated with Gadget-4 \citep{Springel21} modified to include a collisional star cluster \citep{CarlbergK21}. Here we have included stellar \lq\lq{}kicks\rq\rq{} \citep{Meiron21} to mimic binary encounters,  modeled as the addition of  Gaussian random velocities of $\sigma\sqrt{3/2}$ added to each component of the velocity for a random fraction of 0.001 of the stars within half of the virial radius, where $\sigma$ is the 1D velocity dispersion. The cluster expands with time, so we start the simulation with  half mass radius of 5.0 pc. The cluster expands to a half-mass radius of 6.6 pc at 2.29 Gyr although this makes little difference to the dynamics beyond 10 pc.

The dark halo is modeled as a Hernquist sphere \citep{Hernquist90} specified with a mass and scale radius with values motivated by the sub-galactic halos found in our cosmological simulations \citep{CarlbergK21}. The pre-infall mass of the entire Sagittarius galaxy to its virial radius may have been quite large, $1.6\times 10^{10} M_\sun$ \citep{Lokas10} but we are interested in the mass remaining on the current orbit. The dark matter mass within 10 pc of the globular cluster center ( about two half-mass radii) is observationally constrained to be well below the stellar mass, otherwise the velocity dispersion of the cluster stars would be larger than observed.  The dark matter mass within 100 pc must rise to be at least the mass of the stellar cluster in order to have any dynamical effect. We run models with $M_h$= $1\times 10^8 M_\sun$ and $3\times 10^8 M_\sun$, both $a=0.4$ kpc, and $M_h=7\times 10^8 M_\sun$, $a=0.6$ kpc. The star particles have mass 20 $ M_\sun$ and the dark matter particles 40 $ M_\sun$ with softenings of 2 and 5 pc, respectively, which is sufficient to provide an accurate star cluster model at the half mass radius and beyond. The current location and velocity of M54 \citep{Baumgardtetal19} is used to integrate the orbit backward to the orbital apocenter around 2.29 Gyr ago \citep{VB20:Sgr}. 

The star cluster is placed at the center of the dark halo. It is possible that the cluster formed at some distance from the center of the dark halo and that dynamical friction subsequently dragged it inwards. Dynamical friction would have caused the massive, dense, M54 to spiral to the center of the dark matter halo with minimal tidal mass loss. For instance, a straightforward simulation shows that if started on a circular orbit at 0.5 kpc, the star cluster spirals into the center in less than 0.5 Gyr, heating the central dark matter particles to create a 30 pc core in their distribution.

The star cluster and its dark halo adjust quickly to the new equilibrium with each other and their particle gravitational softening (2 and 5 pc, respectively). In the first 0.01 Gyr the star cluster half mass radius increases about 10\% and the dark halo density at 10 pc approximately doubles, with its velocity dispersion going up about 50\%. The simulation responds dynamically on the radial orbit timescale, 0.86 Gyr. Successive pericenter passages heat the outer dark matter halo, which expands over the orbital period and is pulled away at the next pericenter passage.  Tides and internal dynamical heating  of stars  moves about 1\% of the cluster mass into the region between 30 and 200 pc. If there were no dark halo these stars would be unbound from the star cluster and would form a thin tidal stream of stars. Internal relaxation is very slow, as expected for M54 whose relaxation time has been estimated to be about 6 Gyr  \citep{BaumgardtHilker18}. By comparison, our  model  has a relaxation time of 6.3 Gyr.

\begin{figure}
\begin{center}
\includegraphics[scale=0.45,trim=20 30 40 45, clip=true]{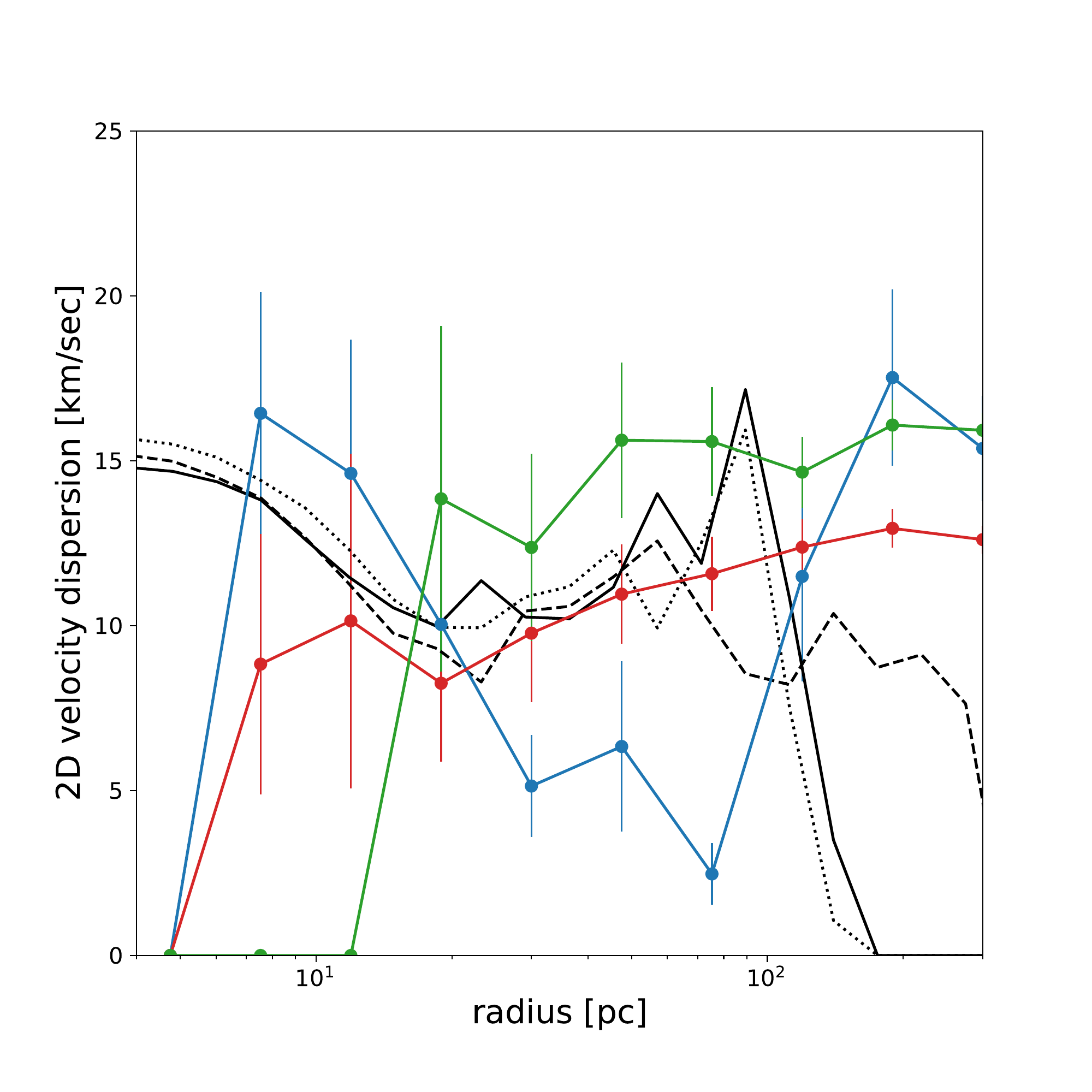}

\includegraphics[scale=0.45,trim=20 30 40 45, clip=true]{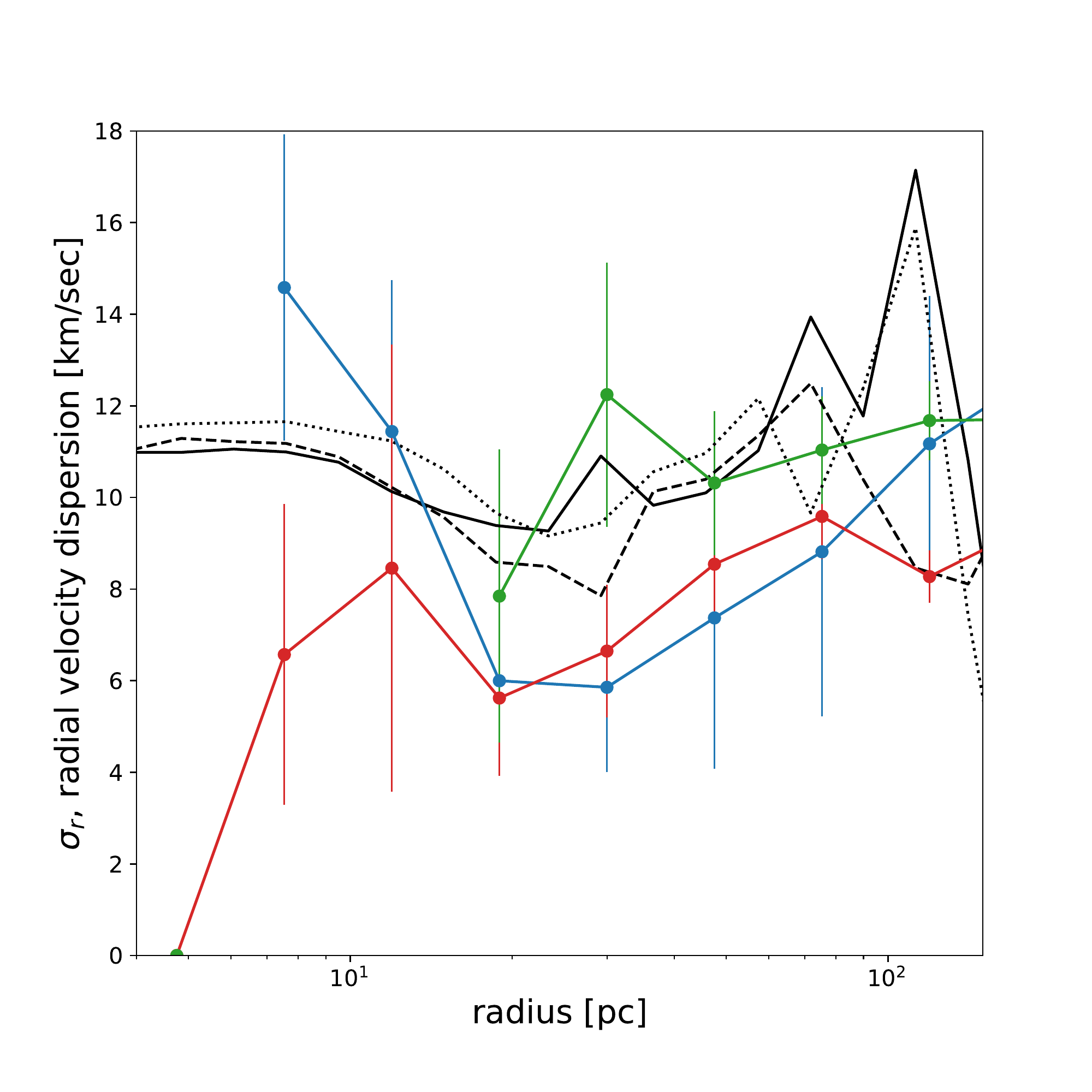}
\end{center}
\caption{The projected 2D (top) and radial (bottom) velocity dispersion of M54\rq{}s  stars compared to the same quantity measured for the star particles in the simulations that start with 1, 3, and 7$\times 10^8$  dark halos, shown as black dashed, dotted and solid lines, respectively.   }
\label{fig_M54sigstar}
\end{figure}

\begin{figure}
\begin{center}
\includegraphics[scale=0.45,trim=0 20 40 45, clip=true]{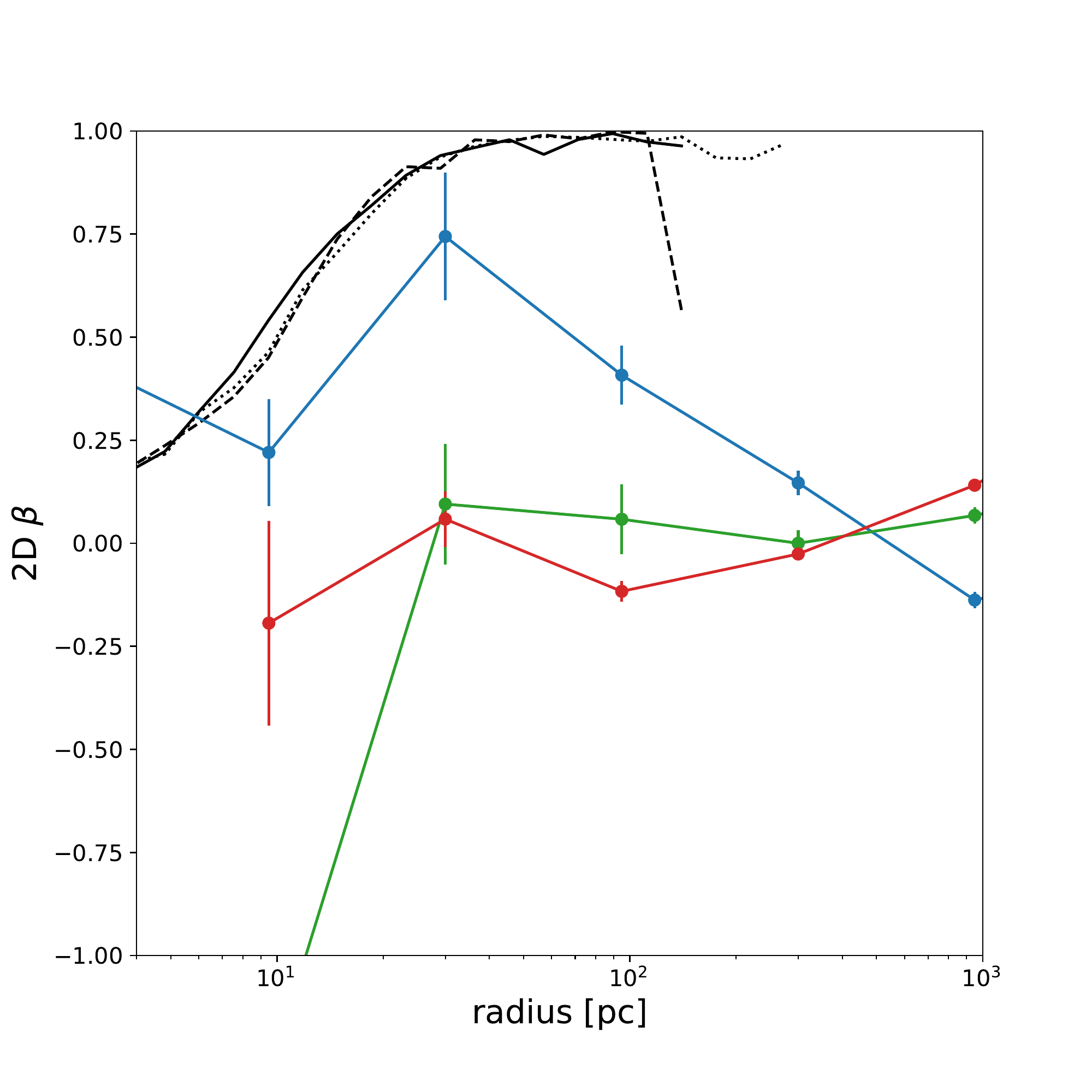}
\end{center}
\caption{The 2D velocity anisotropy, $\beta_{2D} = 1-\sigma_t^2/\sigma_r^2$, for the color subsamples compared to the simulated star cluster's initial dark halo. The velocity dispersions have been corrected for the proper motion velocity errors. The values in the 1, 3, and 7$\times 10^8$  dark halos are shown as black dashed, dotted and solid lines, respectively.}
\label{fig_M54beta}
\end{figure}

\subsection{Evolution of the Simulation}

The model runs for 2.29 Gyr to reach a Galactic position close to that of the observed cluster.  The star particle positions and velocities are projected onto the sky and analyzed in the same was as the M54 cluster stars. The 2D velocity dispersion of M54\rq{}s blue stars are compared to the star particles in the model in Figure~\ref{fig_M54sigstar}. The models predict a velocity minimum around 30 pc and then a rise back to nearly the central velocity dispersion at 100 pc. Our blue subsample shows the same trend. 

The model star cluster particles gradually extend to somewhat more than 100 pc.  The M54/Sgr stars selected to be in the blue subsample extend over the entire surface of the dwarf galaxy, with cluster stars in the center and dwarf galaxy stars at larger radii. A kinematic discriminant between cluster and Galaxy stars is the velocity anisotropy.  Defining $\beta_{2D} =  1-\sigma_t^2/\sigma_r^2$ the model values are compared to the color subsamples in Figure~\ref{fig_M54beta}. The cluster stars in the model are expected to be radially anisotropic in the 30 to 100 pc range, as their orbits are generally perturbed only in the central region of the cluster and become highly elliptical. The green and red samples are essentially isotropic at all radii.  The M54/Sgr data show that the blue sample is radially anisotropic in the range 30-100pc, becoming isotropic at larger radii.  The change in velocity anisotropy with radius is also consistent with the change in density profile shown in Figure~\ref{fig_denr}, with stars inside of 100 pc largely belonging to the cluster while stars beyond 100 pc largely belong to the dwarf galaxy.

\section{The Mean Velocity Field of the Sagittarius Dwarf}

The Sagittarius Dwarf galaxy is being shredded in the Milky Way\rq{}s gravitational field \citep{Ibata94} to produce a tidal stream wrapped around the galaxy at least twice \citep{Ibata20:SgrStream}. Our interest is to use the velocity profile of the inner few kiloparsecs of  Sagittarius stars to estimate the density profile of the dark matter that remains bound to Sagittarius and place some limits on the pre-accretion mass of the system and how it is related to the M54 cluster. The stars in this region have motions that transition from bound to flowing out into the stream. 
The velocity field of the Sagittarius Dwarf has been extensively studied to understand the shape of the stellar system, the total mass profile and the velocity at which stars are fed into the stellar stream \citep{LawMajewski10,Penarrubia10,Lokas10,Frinchaboy12,VB20:Sgr,delPino21,Ramos21}. A point of controversy is the amount of rotation in the inner 0.5-2 kpc of the system. 

\begin{figure}
\begin{center}
\includegraphics[scale=0.5,trim=20 30 0 40, clip=true]{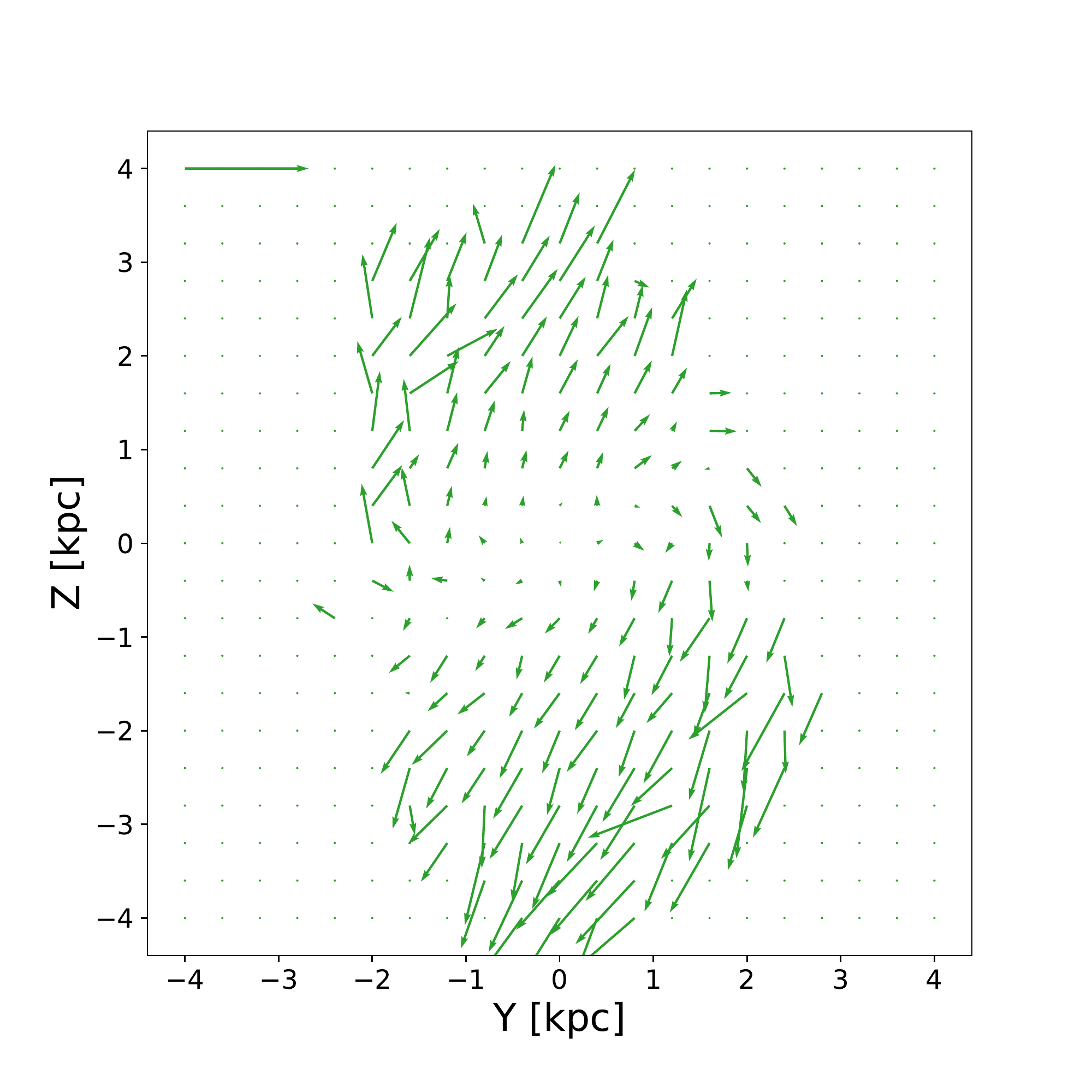}
\includegraphics[scale=0.35,trim=0 0 0 0, clip=true]{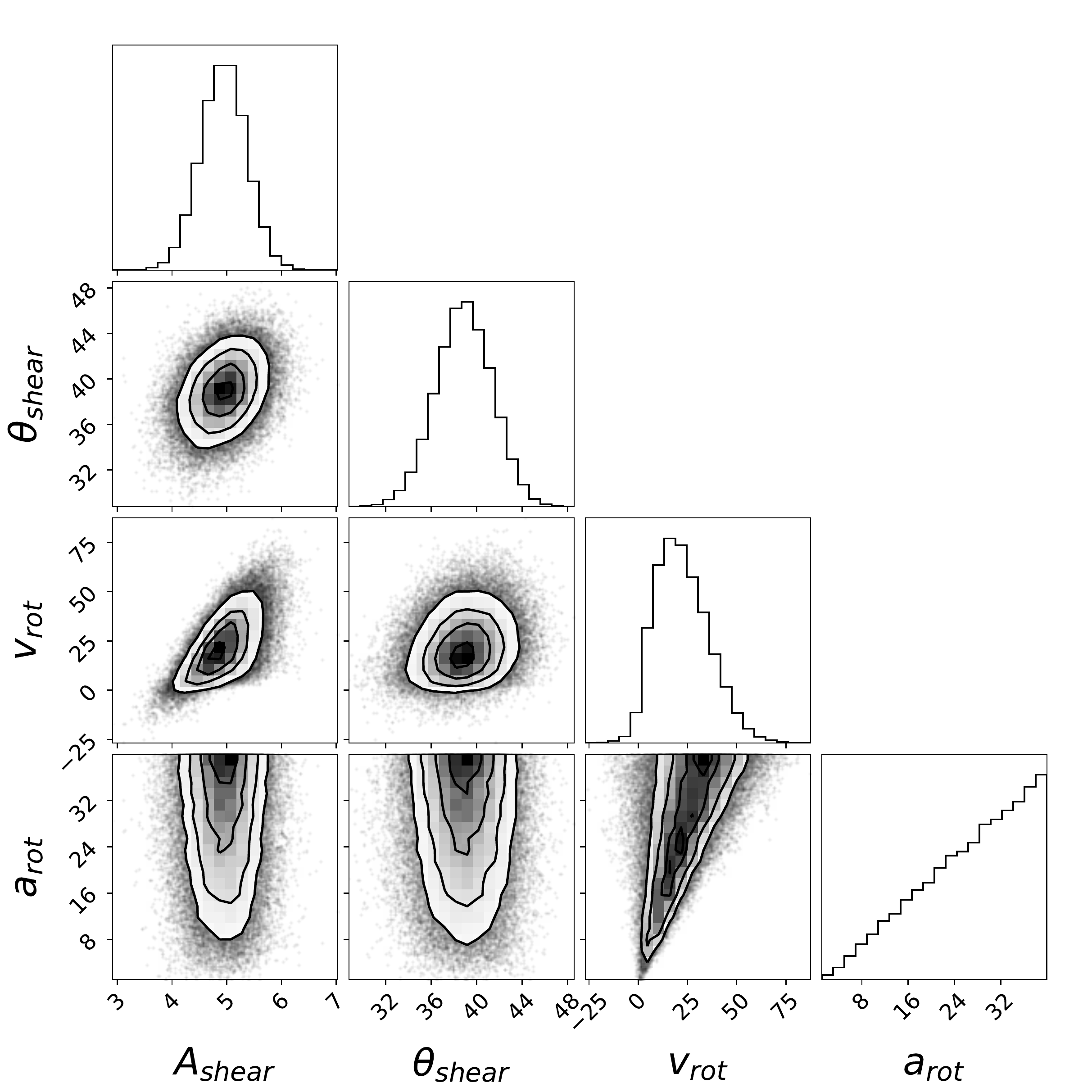}
\end{center}
\caption{The green sample mean velocity field (top panel) and the corner plot of the fit to a rotation plus shear model.
The model removes 73\% of the variance and has a $\chi^2/{\rm dof} = 2.06$}
\label{fig_cornerg}
\end{figure}

\begin{figure}
\begin{center}
\includegraphics[scale=0.5,trim=20 30 0 40, clip=true]{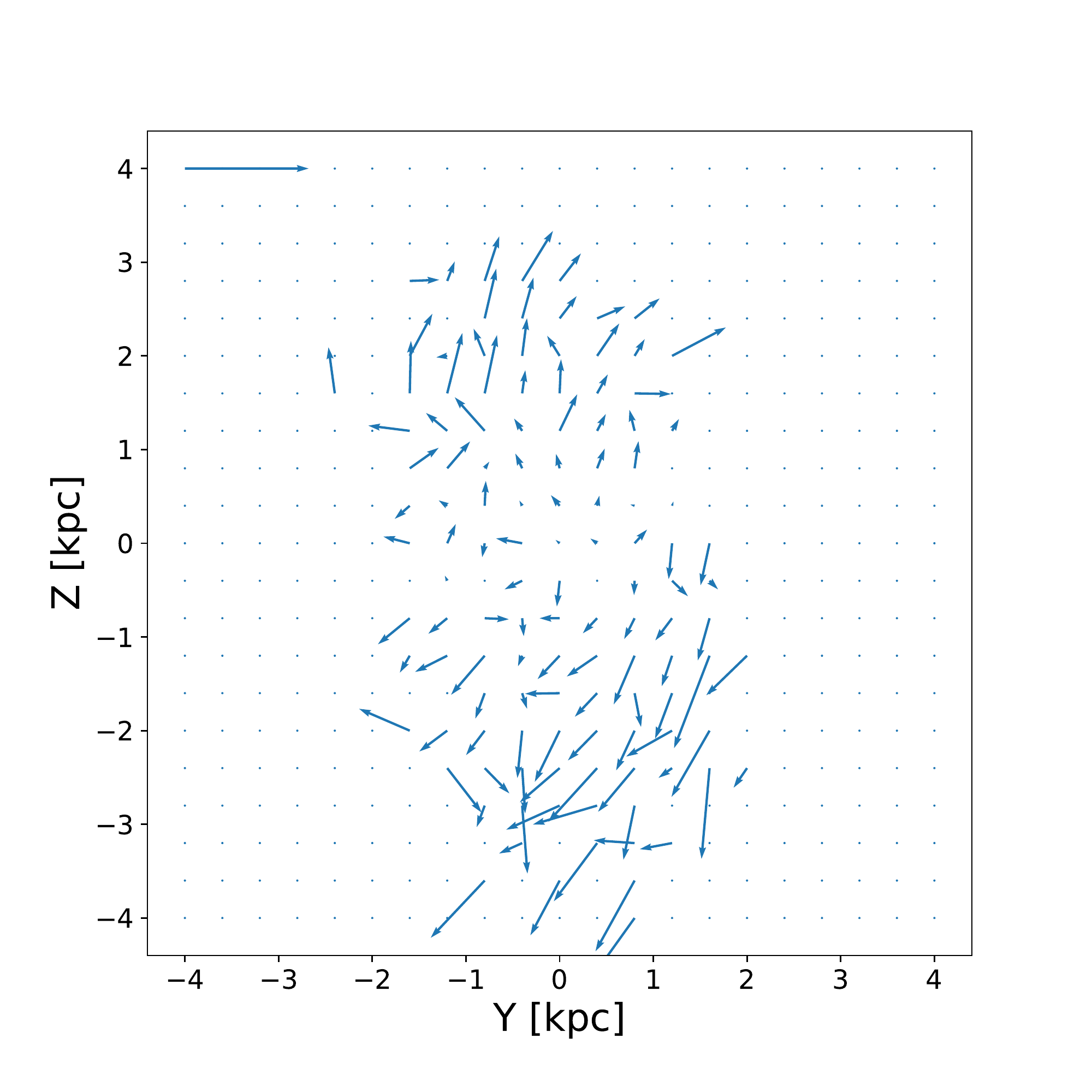}
\includegraphics[scale=0.35,trim=0 0 0 0, clip=true]{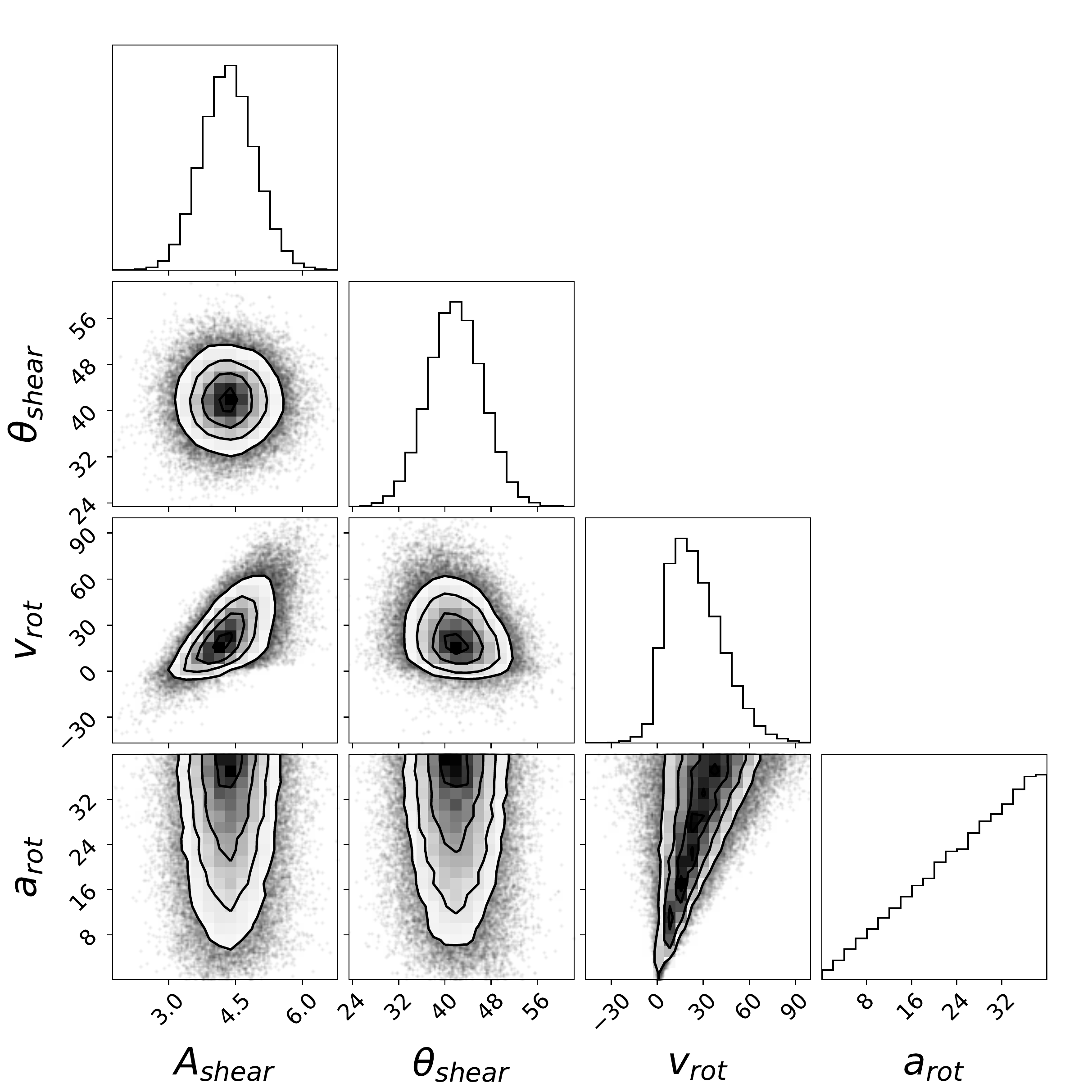}
\end{center}
\caption{The blue sample mean velocity field (top panel) and the corner plot of the fit to a rotation plus shear model.
The model removes 55\% of the variance. $\chi^2/{\rm dof} = 1.10$}
\label{fig_cornerb}
\end{figure}

\begin{figure}
\begin{center}
\includegraphics[scale=0.5,trim=20 30 0 40, clip=true]{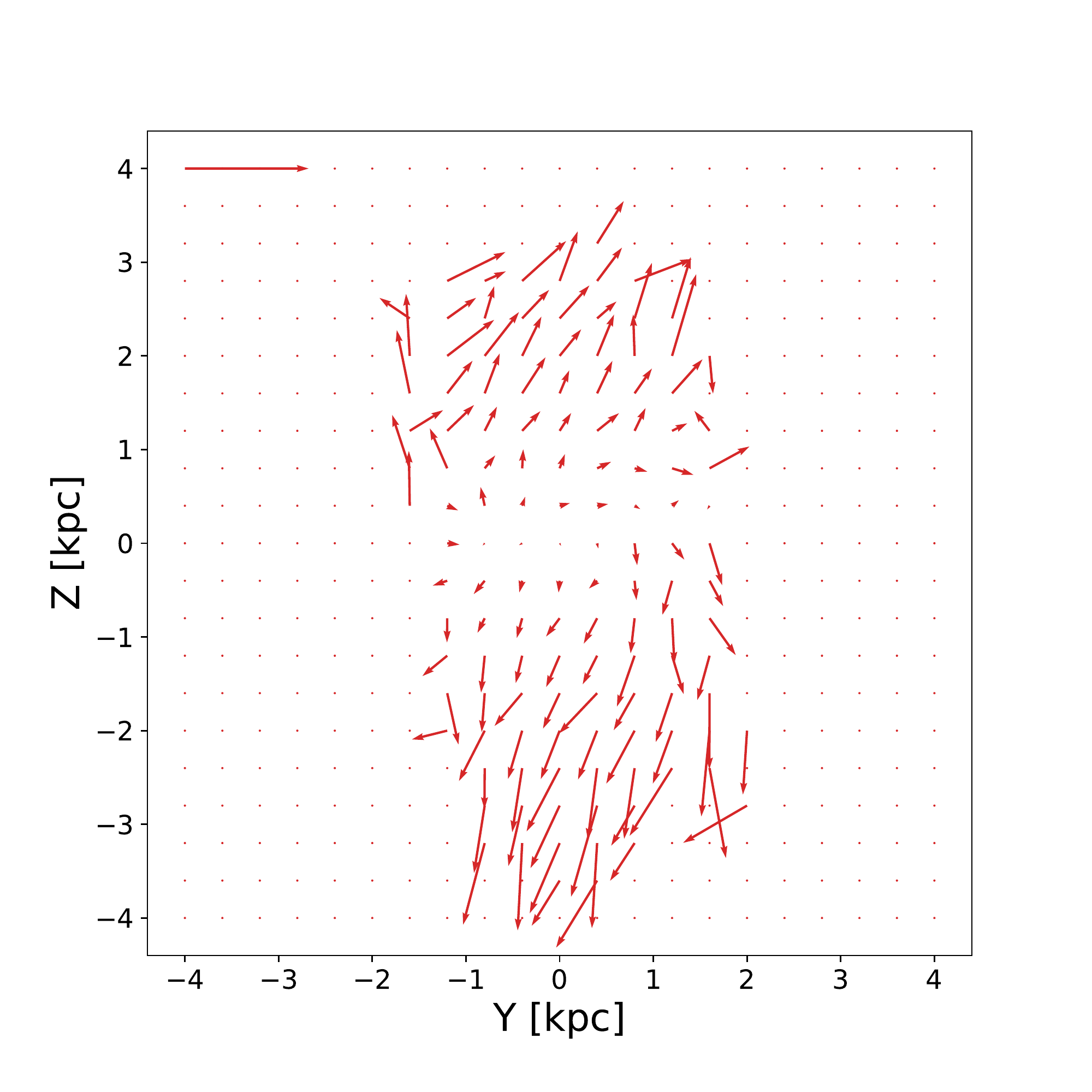}
\includegraphics[scale=0.35,trim=0 0 0 0, clip=true]{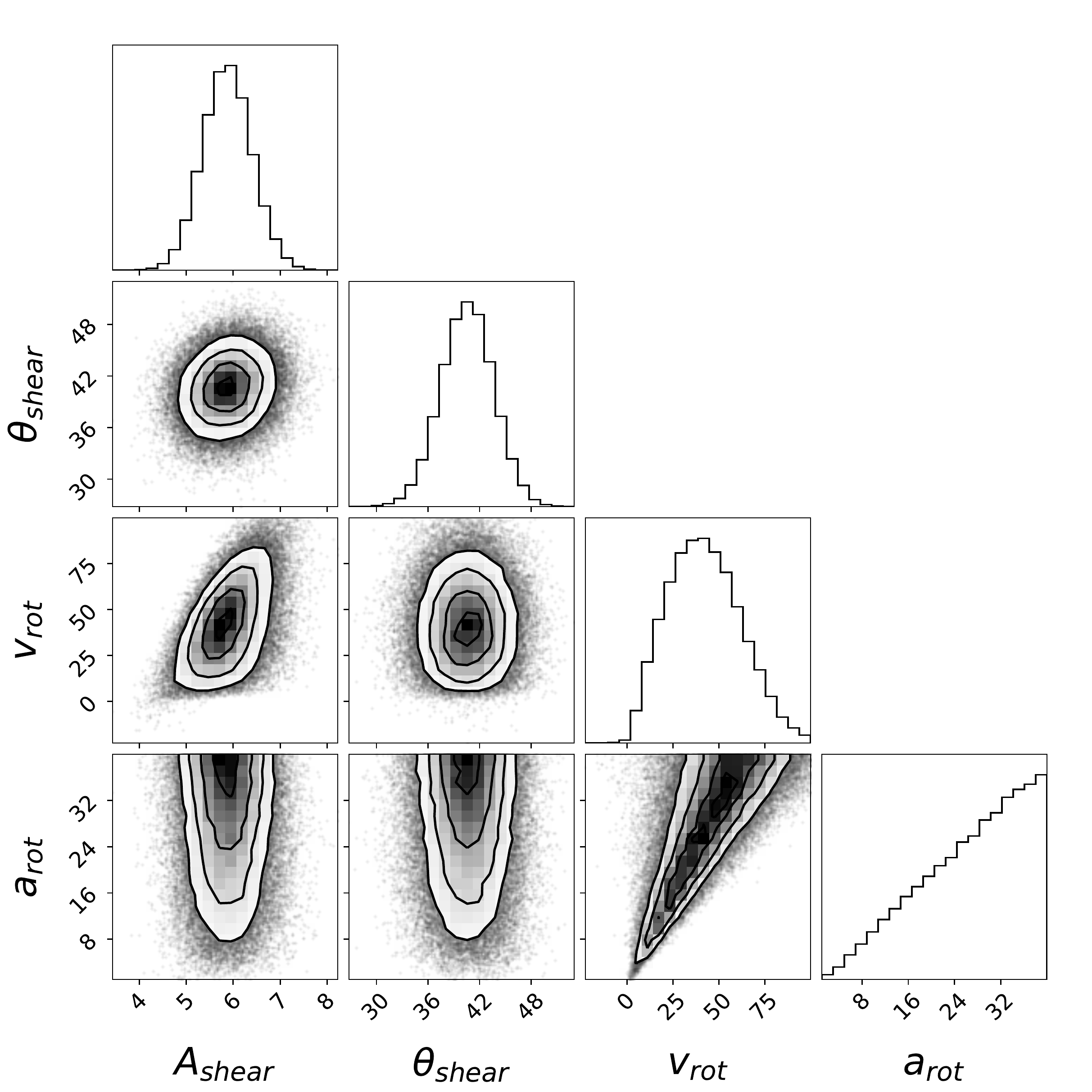}
\end{center}
\caption{The red sample mean velocity field (top panel) and the corner plot of the fit to a rotation plus shear model.
The model removes 66\% of the variance. $\chi^2/{\rm dof} = 2.02$}
\label{fig_cornerr}
\end{figure}

\subsection{The Sagittarius Galaxy Velocities \label{sec_sgr}}

Our color subsamples are separately binned into 0.4 kpc square pixels in a 8 kpc frame centered on the M54 cluster. The mean velocity fields on the plane of the sky (YZ Galactic coordinates relative to the cluster center) are shown in the upper panels of Figures \ref{fig_cornerg}, \ref{fig_cornerb} and \ref{fig_cornerr}. The motions are modeled as the sum of  a rotation and a shear, 
\begin{eqnarray}
v_t(r) &=& v_{rot} {r \over {\sqrt{a^2+r^2}}}, \\
v_r(r) &=& A_r r \\
\vec {v}_{s}(r,\theta) &=& A_s r [\cos{(\theta-\theta_0)},\sin{(\theta-\theta_0)}], 
\end{eqnarray}
which are projected into the x and y components of the velocity. The parameters of the model are $A_r$ the linear expansion rate, $v_{rot}$ the peak of the maximum of the rotation velocity, $a$ the scale radius of the rotation velocity, $A_s$ the velocity shear, and $\theta_0$ the direction of the shear.  The model is fit to the data using the Monte Carlo Markov Chain algorithm emcee \citep{emcee}. The logarithm of the likelihood function is $-\onehalf\sum [\Delta(v_x)^2 + \Delta(v_y)^2]/\sigma_v^2  + \log{(\sigma_v^2)}$, where $\Delta(v_x)$ is the difference between the observed $v_x$ and the value that the model predicts. The $\sigma_v$ is taken from the measured 2D velocity dispersion which is essentially uniform beyond 0.2 kpc from the cluster, nearly independent of color. We set $\sigma_v$ to be constant at 20 \kms. The priors on the model quantities are set to be flat with such a large range that the results have  no dependence on the range of the prior.

The resulting corner plots showing the fits are displayed in the lower panels of Figures \ref{fig_cornerg}, \ref{fig_cornerb} and \ref{fig_cornerr}. The model accounts for 44 to 61\% of the variance in the data overall so the fits are reasonable, with $\chi^2/{\rm dof}$ values ranging from 1.1 to 2.1 assuming a random velocity of 20 \kms\ in each component.  Nevertheless, the model does indicate that all three subsamples have similar mean motions, with an expansion of $A_r=1.6-2.0 \kms\, {\rm kpc}^{-1}$ and a  shear of $A_s = 2-2.5 \kms\, {\rm kpc}^{-1}$,  at an angle of 40-45\degr\  with respect to the Galactic plane.

\subsection{An Artificial Star Dwarf Galaxy \label{sec_dwarf}}

\begin{figure}
\begin{center}
\includegraphics[scale=0.45,trim=0 30 20 30, clip=true]{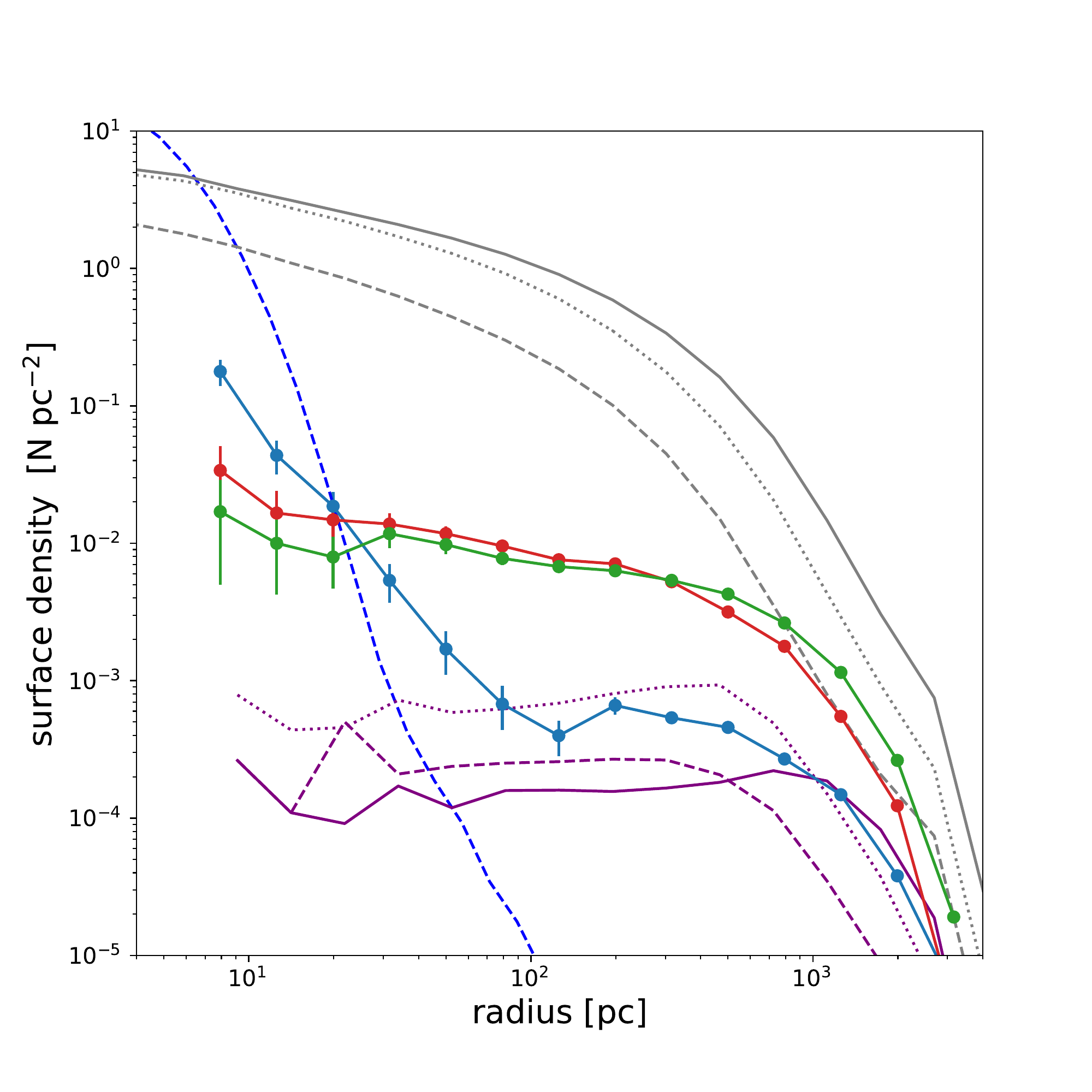}
\end{center}
\caption{The M54-Sgr stellar surface densities with radius (blue, green, red), the n-body star cluster surface density (blue dashed), the artificial dwarf galaxy star profiles (purple) and the dark matter surface density in units of $10^3\,{\rm M}_\sun {\rm pc}^{-2}$ (gray), where the  1, 3 and $7\times 10^8 M_\sun$ simulations have dashed, dotted and solid lines, respectively.}
\label{fig_denrsd}
\end{figure}

\begin{figure}
\begin{center}
\includegraphics[scale=0.45,trim=0 30 20 30, clip=true]{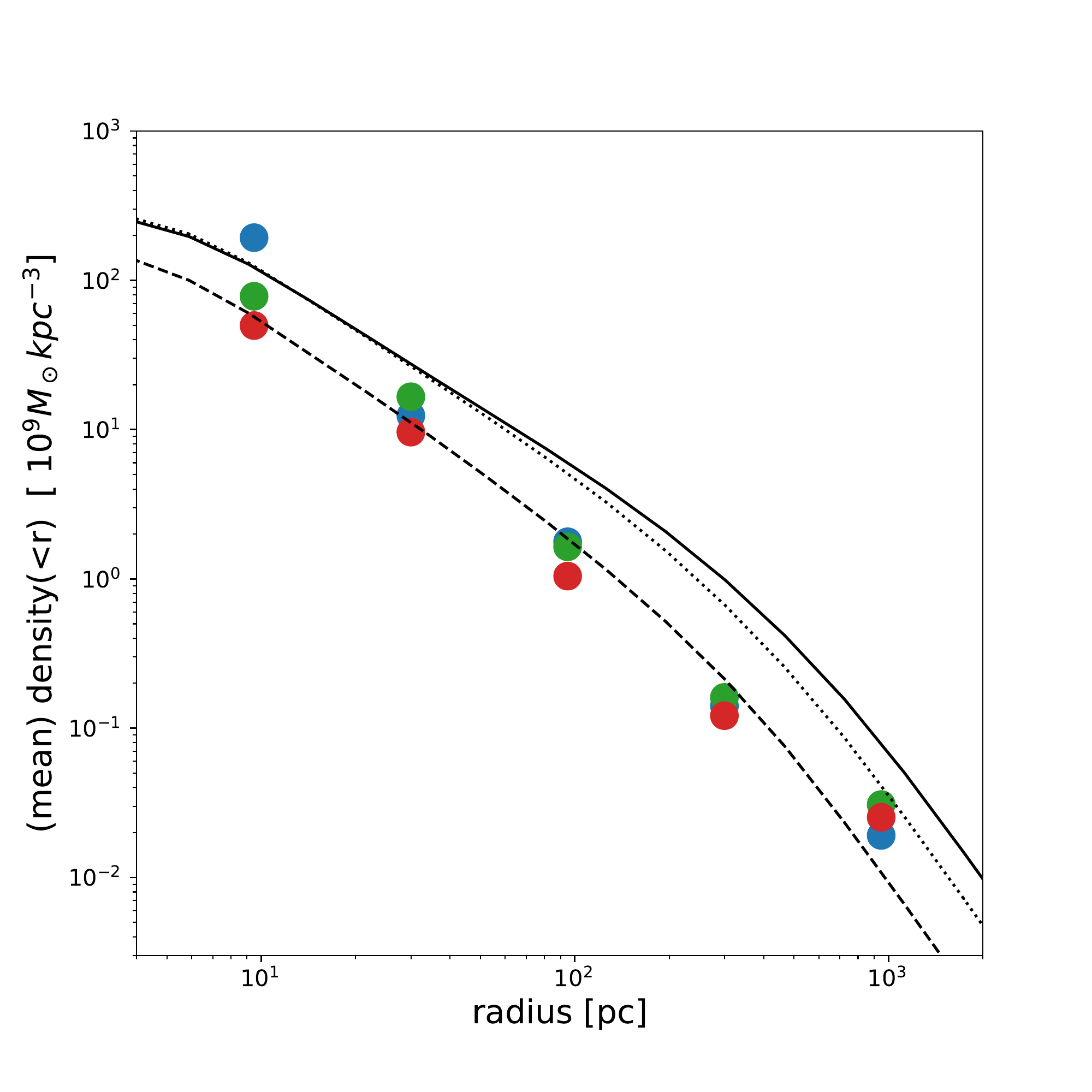}
\end{center}
\caption{The mean 3D density inside radius $r$  inferred from the Jeans equation in Figure~\ref{fig_jeansmass} and the 3D dark matter densities of the  $3$ and $7\times 10^8 M_\sun$ simulations as dotted and solid lines, respectively. The solid grey line is the local 3D density in the  $7\times 10^8 M_\sun$ simulation. All simulation measurements are at 2.29 Gyr.  }
\label{fig_den3d}
\end{figure}

\begin{figure}
\begin{center}
\includegraphics[scale=0.45,trim=5 30 0 40, clip=true]{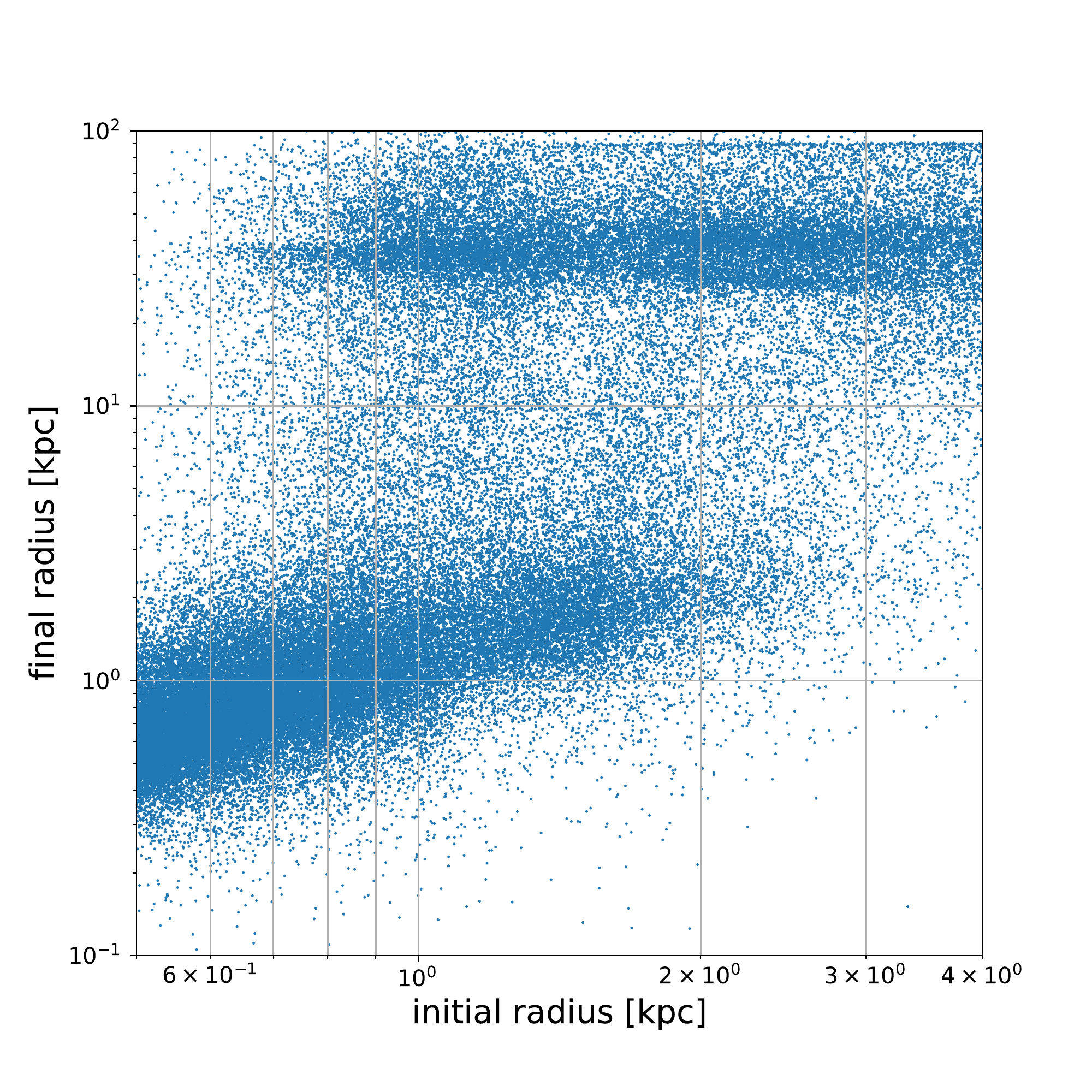}
\end{center}
\caption{The starting and final radii of the artificial dwarf galaxy stars for the $3\times 10^8 M_\sun$ simulation. No particles are selected as stars inside 0.5 kpc, but the particles are on sufficiently elliptical orbits that over time the central hole fills. }
\label{fig_rr}
\end{figure}

\begin{figure}
\begin{center}
\includegraphics[scale=0.44,trim=20 0 0 40, clip=true]{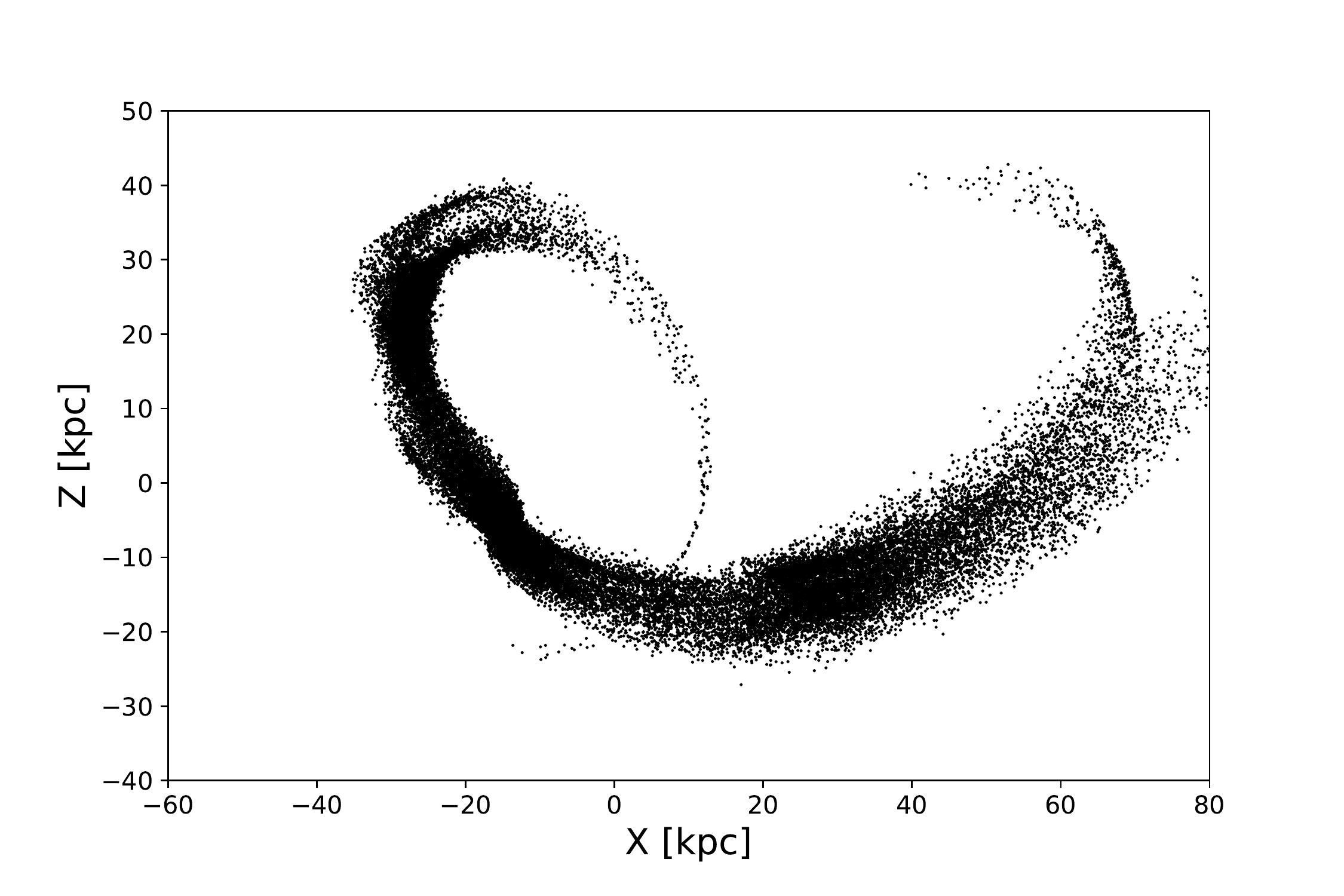}
\end{center}
\caption{The stream that the artificial dwarf galaxy star particles produce in the $3\times 10^8 M_\sun$ simulation. The Galactic center is located at (0,0) and the star cluster is located close to the observed position of M54, at Galactic coordinates [-14.7, 2.2,-6.2] kpc. The simulation began at the orbital apocenter 2.29 Gyr in the past.  A dwarf galaxy rotating in the opposite direction does not produce a bifurcated stream.  }
\label{fig_XZ427}
\end{figure}

To approximate the kinematics of  the stars of the Sagittarius dwarf we select a thick disk-like set of particles from the simulation at early times.
We select from the simulation near the outset, at an age of 0.05 Gyr, identifying a subset of the dark matter particles around the star cluster which are in a flattened distribution, tilted to the line of sight. Sagittarius is a low surface brightness galaxy in which the stars provide relatively little mass compared to the dark matter \citep{Ibata97,Mateo98}. The model dwarf galaxy is constructed 2.24 Gyr ago shortly after the system passed apocenter. The initial disk orientation is taken to be the same as the current orientation, with a unit axis [-0.13, 0.96, 0.26] \citep{delPino21}. The artificial stars are selected as dark matter particles within 4 kpc of the center in the plane of the disk, with a height above the disk less than 0.3 of their radius and a vertical velocity of 15 \kms or less. The 0.05 Gyr time mass profile of the 1-3 and $7\times 10^8 M_\sun$ halos are reasonably approximated with Hernquist spheres with scale radii of 0.35 and 0.55 kpc, respectively, slightly smaller scale radii than the initial values. The dark matter halo mass model defines a circular velocity with radius and the energy and angular momentum of circular orbits with radius. At 1 kpc the circular velocity in the halo is 23 \kms. Particles with energies between 0.8 and 1.2 of the circular velocity energy and 0.8 to 1.2 of the circular velocity energy within  $r_{max}$ of 4.0 kpc are selected at an initial time of 0.05 Gyr as an artificial first approximation to the distribution of dwarf galaxy stars. To create a flat density profile we select particles with $r_{min}>$  0.5 (1.0) kpc, for the  3 and $7\times 10^8 M_\sun$ models, respectively. At the time of selection the particles are in a tilted, rotating disk of 4.0 kpc extent and an aspect ratio of about 1:3. The mean rotation of the stars is counterclockwise around the angular momentum vector,  but clockwise when viewed in the plane of the sky.

\begin{figure}
\begin{center}
\includegraphics[scale=0.45,trim=5 30 0 40, clip=true]{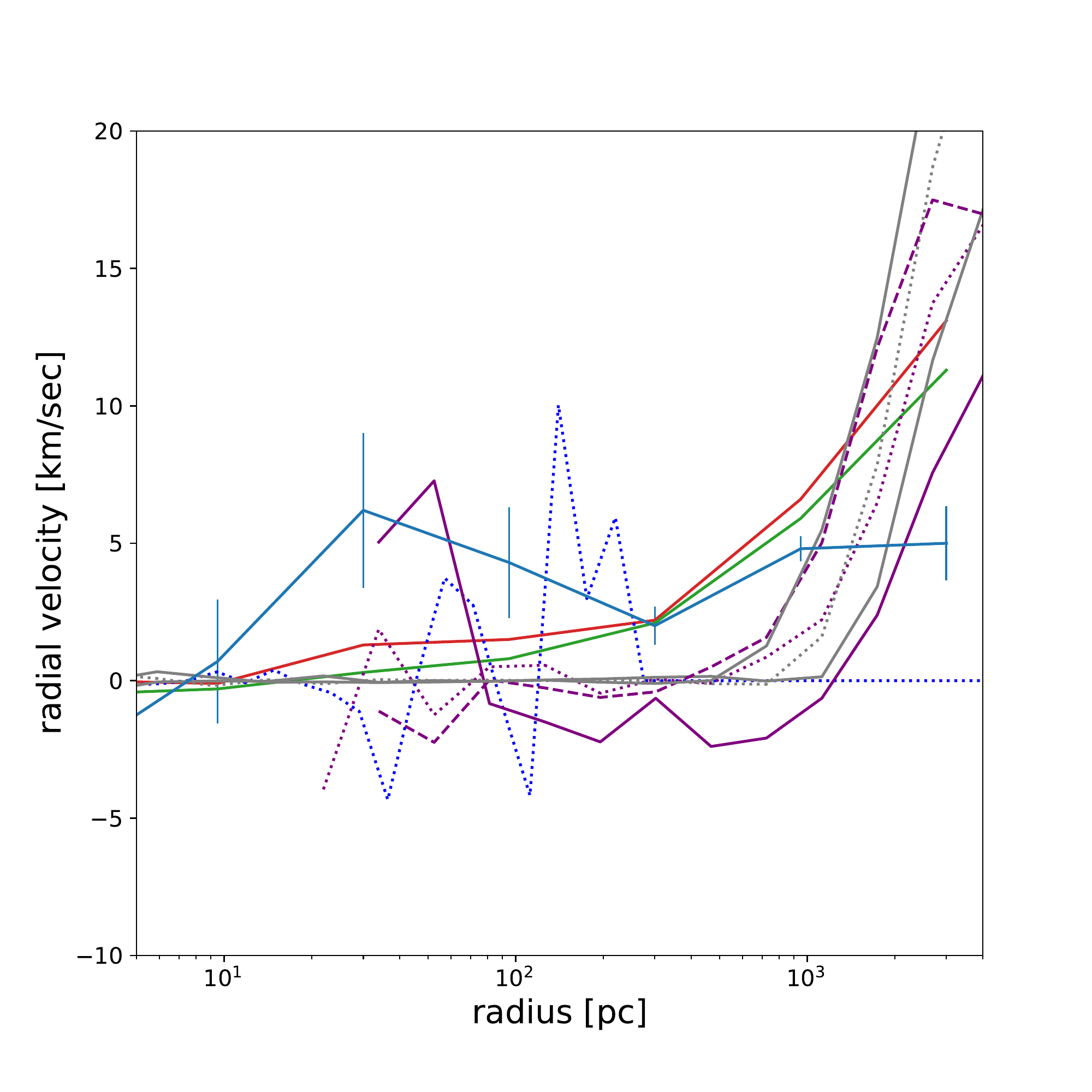}
\includegraphics[scale=0.45,trim=5 30 0 60, clip=true]{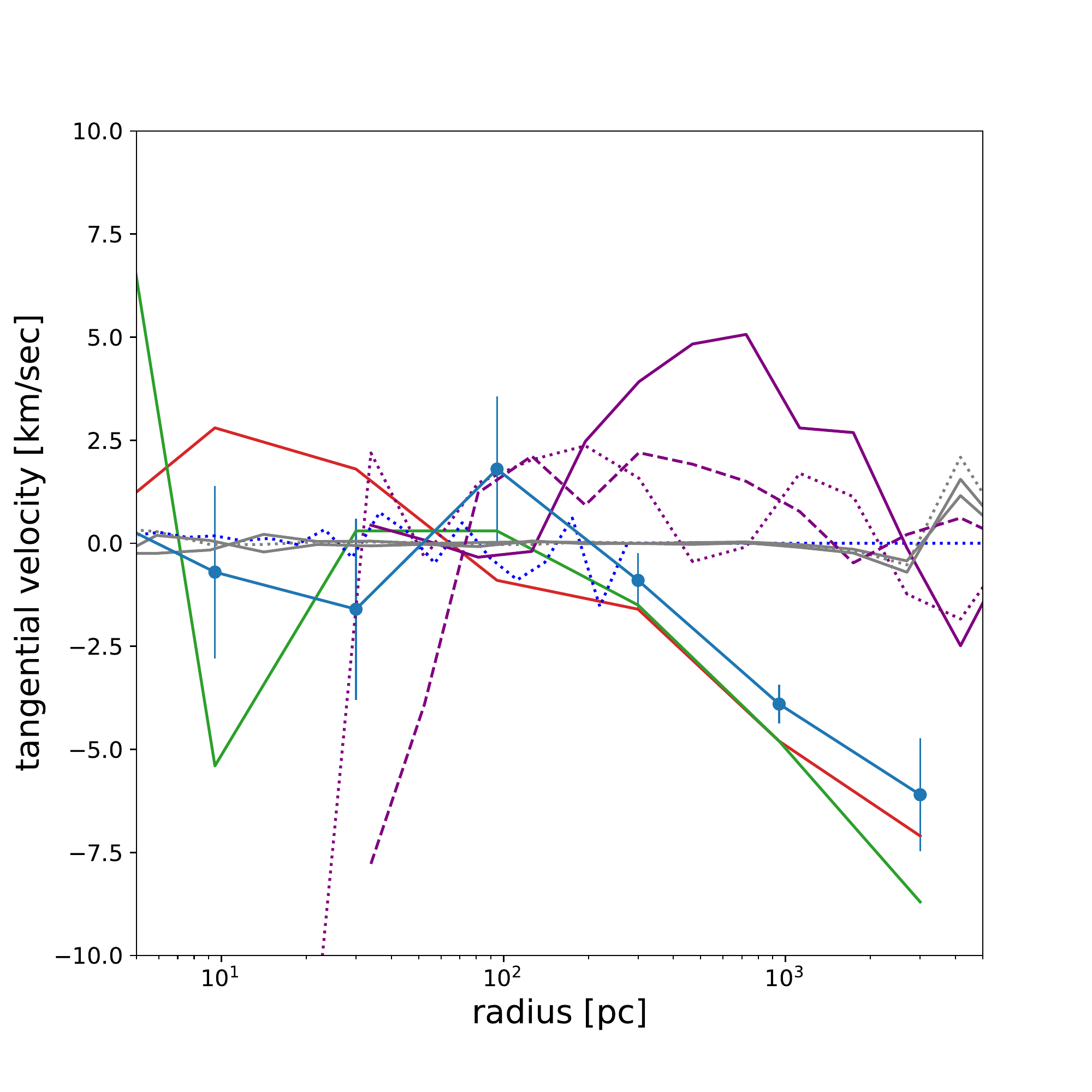}
\end{center}
\caption{The radial (top) and tangential (bottom) velocities with radius for the M54/Sgr data and the artificial stars, selected with positive plane-of-the-sky tangential velocities. The central model cluster star particles are shown as dotted blue lines. The other lines are as in Figure~\ref{fig_denrsd}.}
\label{fig_vrt}
\end{figure}

\begin{figure}
\begin{center}
\includegraphics[scale=0.5,trim=20 30 0 40, clip=true]{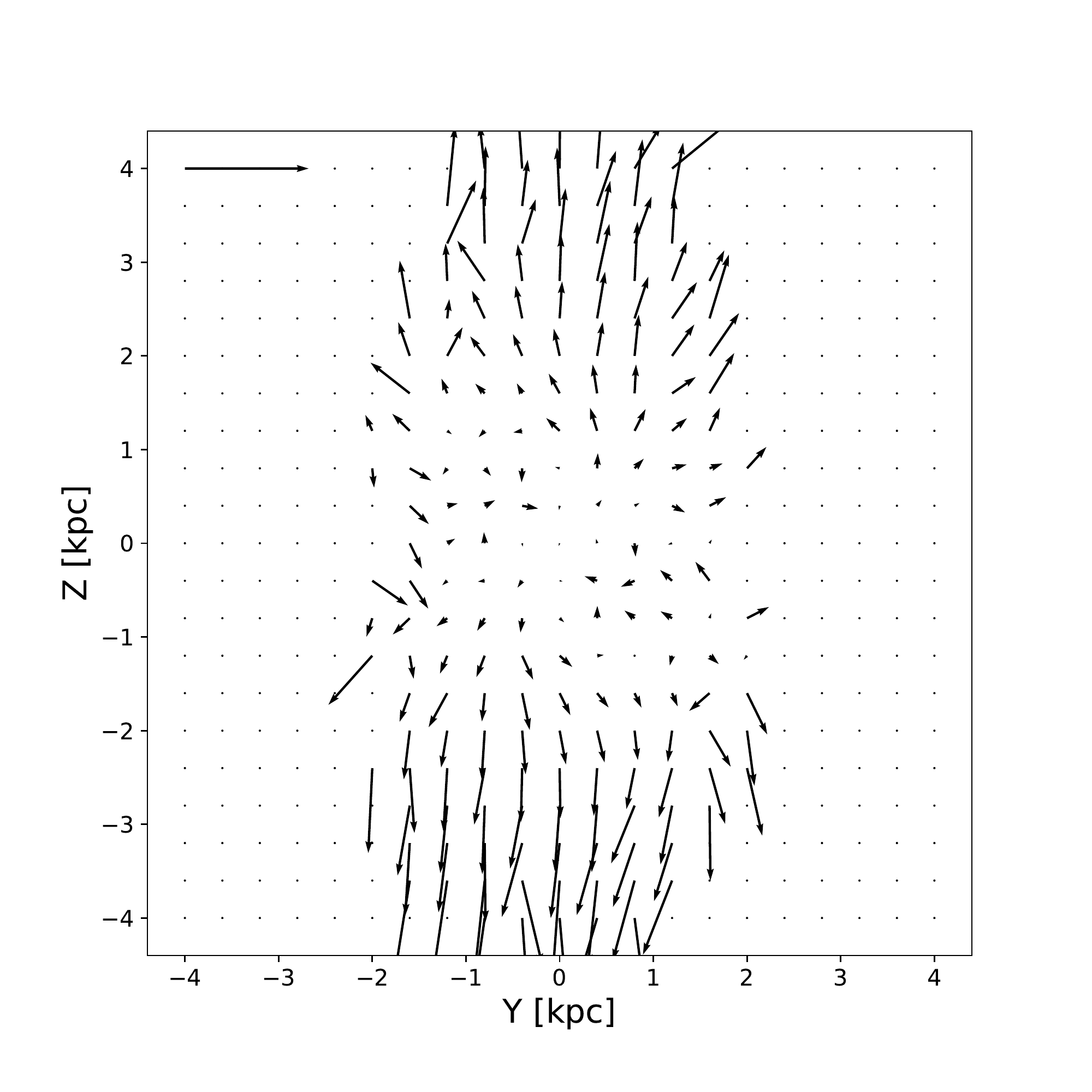}
\end{center}
\caption{The mean velocity vectors in the plane of the sky of the artificial dwarf galaxy particles in the  $3\times 10^8 M_\sun$ simulation. The plot is is comparable to velocity map panels of Figures~\ref{fig_cornerg}, \ref{fig_cornerb} and \ref{fig_cornerr}. The arrow in the upper left has a length of 30 \kms.}
\label{fig_qn}
\end{figure}

The projected surface density profile of the n-body stellar cluster (dashed blue line) is shown in Figure~\ref{fig_denrsd}. The comparison with the density profile of the old, metal-poor stars of our Gaia subsample illustrates the tremendous undersampling of stars in the central region of the star cluster and how the stars of the cluster overlap with the old, metal poor, stars of the dwarf galaxy beyond 50 pc. Three halo models, with initial masses of 1, 3 and 7$\times 10^8 M_\sun$, were evolved along with the star cluster, with the projected surface density of the dark matter halos at the time that matches the current orbital location are shown as the dashed, dotted and solid gray lines, respectively. Randomly selecting particles from the entire dark matter halo will not mimic the Sgr dwarf density profile, which requires that the fraction of dark matter particles identified as stars decline inward about a factor of 100 from 1 kpc to 10 pc. The simple identification procedure here leads to the surface densities shown as purple lines in Figure~\ref{fig_denrsd}. The procedure with an initial central hole in the selected stars leads to a somewhat too flat density profile which could be fixed by allowing some of the stars in the central region. The main outcome is that the dwarf galaxy stars are a decreasing fraction of the dark matter towards its center.

In Figure~\ref{fig_den3d} the mean interior density, $\overline{\rho}(r) = M(r)/(4\pi/3 r^3)$  of the dark matter (without the star cluster mass) derived from the three independent color subsamples is compared to the three n-body dark matter mean interior halo densities at the current location of the M54-Sgr system. Within this limited range of these three models the initial dark matter halo of $3\times 10^8 M_\sun$ and scale radius 0.4 kpc gives results closest to the observed values.

The distributions of the artificial dwarf galaxy star particles are shown in Figure~\ref{fig_XZ427} for the simulation with a total halo mass of $3\times 10^8 M_\sun$. The Galactic XZ plane is shown since the orbit is nearly in this plane. The stream in the $M_h=7\times 10^8 M_\sun$ model is slightly wider, longer, and fuzzier. The lower mass model  produces a significantly narrower stream. A similar plot, with the direction of the X axis reversed, is shown in Figure 10 of \citet{VB20:Sgr}. Their simulations are based on dark halos with  circular velocity curves that peak in the range 2-4 kpc, whereas here it peaks at around 0.4 kpc.  

It is interesting to note that the selection of positive tangential velocities (projected into the plane of the sky)  leads to  stream bifurcations, as seen in the Sagittarius stream \citep{Belokurov06:FieldofStreams,Koposov12:Bifurcation}.  The same selection with negative velocities does not have prominent bifurcations, confirming the importance of rotation in the dwarf galaxy for their origin \citep{Ramos21}. 

One of the uses of the artificial dwarf galaxy stars is to identify effective tidal radii, within which stars remain bound to the dwarf and beyond which they eventually join the massive tidal stream. The radii of stars relative to the star cluster at 0.05 Gyr and 2.29 Gyr are displayed in Figure~\ref{fig_rr}. The artificial star particles that join the stream have a clearly defined minimum radius, about 0.5 kpc, for $3\times 10^8 M_\sun$ halo. Beyond the minimum radius an increasing fraction of particles join the stream with radius, with almost all joining at 2 kpc for the $3\times 10^8 M_\sun$ halo. The absence of a crisp tidal radius is a consequence of the elliptical Galactic orbit of the system and the eccentricity of the particle orbits.

The velocities and positions of the artificial dwarf galaxy stars are projected onto the sky as seen from the location of the Sun in the Milky Way at time 2.29 Gyr for the two halo simulations. The velocities are measured in the same way as for the observational data. The resulting, radially averaged velocity fields are shown in Figure~\ref{fig_vrt}. Figure~\ref{fig_vrt} shows the mean radial velocities in the top panel and the corresponding model measurements. The positive bump in the measured mean radial velocity of the blue stars has a large uncertainty but is also seen at a somewhat larger radius in the n-body model star cluster particles, where it is likely a result of the Galactic tidal field for the system which passed pericenter 0.04 Gyr before the time being analyzed. Figure~\ref{fig_vrt} shows the mean tangential velocities in the bottom panel. The $3\times 10^8 M_\sun$ model has tangential velocities closer to the observed values over the 100-1000 pc range, leading us to favor it as the better velocity model inside 1 kpc. 

Beyond 1 kpc the strong shear seen in Sgr shows up as the outer tangential velocity decline in Figure~\ref{fig_vrt}, which  is not present in the projection of any of the models.  The artificial disk star velocities plotted over the face of system shown in Figure~\ref{fig_qn} are similar in the inner 1-2 kpc but not a good fit to the velocities of Figures~\ref{fig_cornerg}, \ref{fig_cornerb} and \ref{fig_cornerr} beyond 2 kpc. The model system has a smaller shear velocity from top to bottom than observed in Sagittarius. 

The model system has a large velocity gradient along the line-of-sight velocity at the current orbital phase as can be seen in Figure~\ref{fig_XZ427}, where the segment of the stream below M54 is moving nearly horizontally away from the galactic center, whereas the part of the system at the top, closer to the plane, is turning upward with a reduced line of sight velocity.  To increase the shear in the model one solution would be to rotate the orbit about 5\degr, to project more of the strong velocity shear into the plane of the sky. Another possibility is that the cylindrically symmetric potential used here does not allow for the Galactic bar, although the asymmetry of the potential due to the bar will be small at a pericenter distance of 14 kpc.

\section{Alternative Futures Evolution}

\begin{figure}
\begin{center}
\includegraphics[scale=0.45,trim=10 30 10 40, clip=true]{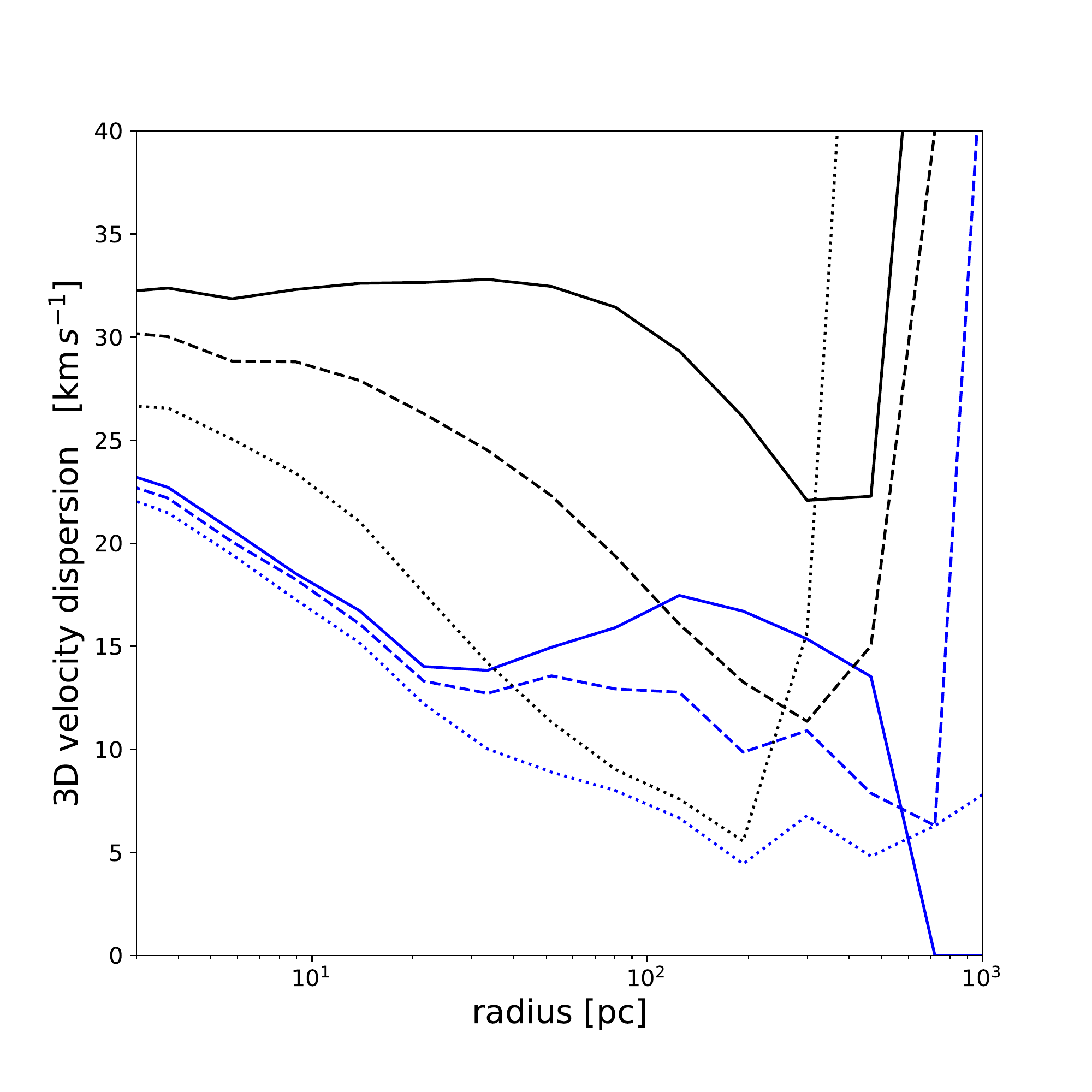}
\includegraphics[scale=0.45,trim=10 30 10 60, clip=true]{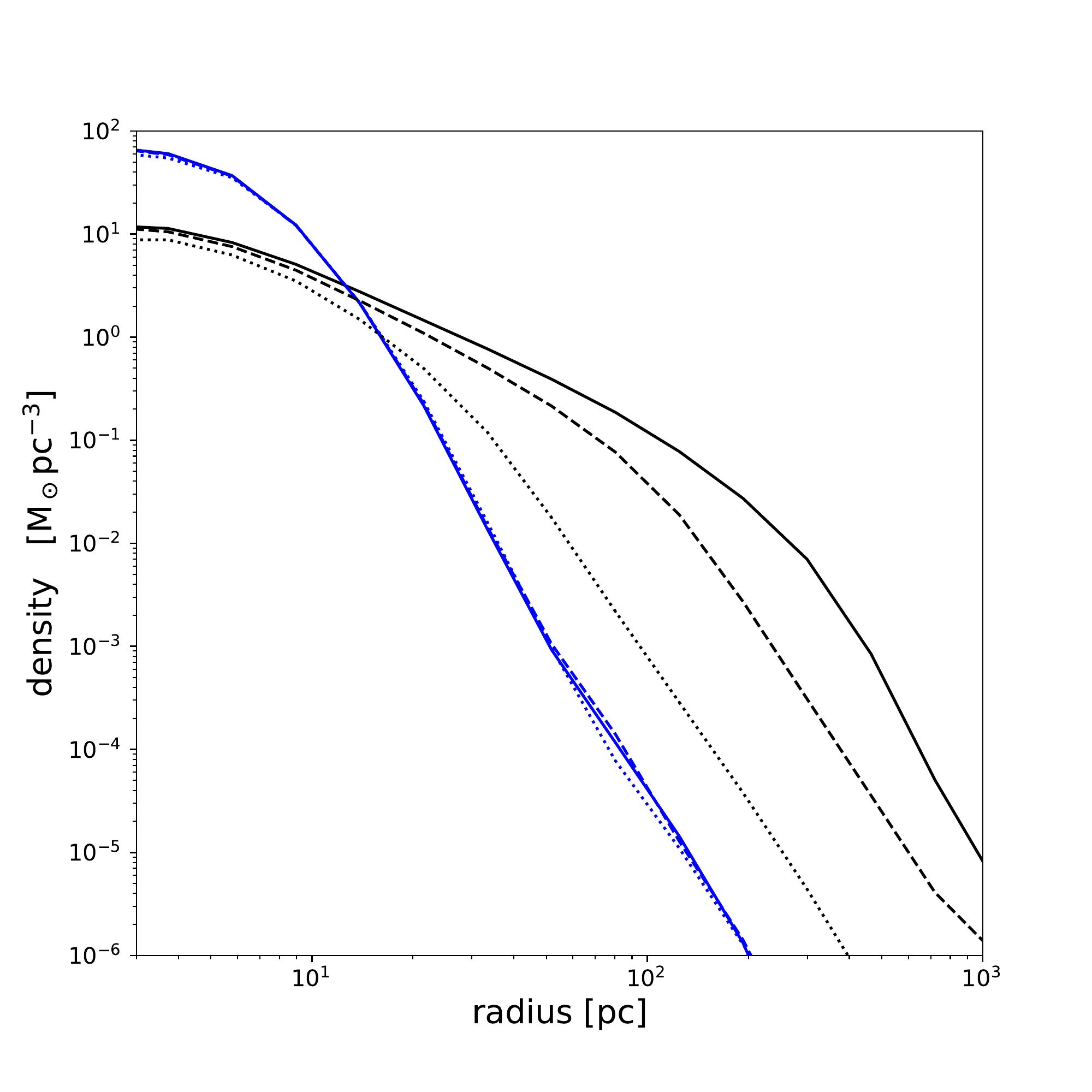}
\end{center}
\caption{Top: The 3D velocity dispersion at 10 Gyr for M54\rq{}s orbit (solid) and the orbit started at apocenter  with 1/2 (dashed) and 1/4 (dotted) of the angular momentum. The black line is for the dark matter, the blue for the cluster stars. Bottom: the density profiles for the same models with the same line designation.}
\label{fig_endpoints}
\end{figure}

Our radial velocity dispersion measurements, their population density slopes, and $\beta$ values lead to the mass estimates shown in Figure~\ref{fig_jeansmass}.  Our Jeans equation mass analysis finds 1.4, 1.9 and 1.1$\times 10^6 M_\sun$ within 30 pc, for the blue,  green and red samples, respectively. The Jeans masses are essentially consistent with the cluster\rq{}s stellar mass of $1.41-1.78\times 10^6 M_\sun$ \citep{BaumgardtHilker18}. At 95 pc the Jeans analysis masses are 6.3, 5.9 and  3.7$\times 10^6 M_\sun$, and at 300 pc the masses are 15.9, 18.2, and 13.8$\times 10^6 M_\sun$, for the blue,  green and red samples, respectively. The implied rise of mass with radius is $r^{1.20}$ between 30 and 95 pc and $r^{1.08}$ between 95 and 300 pc, with a total increase in mass a factor of 10.9 between 30 and 300 pc. The mean density of the dark matter is close to what is expected for a CDM dark matter halo of around $3\times 10^8 M_\sun$ initiated at the apocenter of the frictionless M54 orbit 2.29 Gyr ago. 

The mass and extent of the remnant globular cluster dark matter halo depends on how long ago the system fell into the Milky Way and how close the orbit comes to the center of the galaxy, where the tidal field dynamically heats the system\rq{}s particles and pulls them away in a tidal stream. Over the 2.29 Gyr of evolution the models lose about 16\% of the mass within 1 kpc to tidal fields. It is not unlikely that systems like Sagittarius with a massive globular cluster near their centers fell into the Milky Way much earlier when the bulk of the dark matter mass was built up, which will lead to more tidal mass loss.  Simply running the simulation to 10 Gyr we find that the $3\times 10^8 M_\sun$ halo loses 38\% of its mass inside 1 kpc, but only 9\% inside 0.1 kpc. Had the orbit come closer to the Galactic center even more dark matter mass would have been lost.  Two further simulations start at the same apocenter as used for M54, but reduce the initial orbital velocities so that the system has 1/2 and 1/4 of the angular momentum. The simulations use the same static Galactic potential with no allowance for dynamical friction. The resulting star and dark matter velocity and density  dispersion profiles are shown in Figure~\ref{fig_endpoints}. As expected, smaller pericenter passages remove more dark matter more  quickly. The velocity dispersion profile of the system on the 1/4 angular momentum orbit is always declining to 100 pc.  Tidal fields remove 90\% of the mass inside 1 kpc within the first 2 Gyr. If  gas were present, it is likely that ram pressure would remove much of the gas as well \citep{SgrGas:18}. Therefore, it appears that if M54 had fallen in earlier on a smaller pericenter orbit it would have formed few if any surrounding dwarf galaxy stars and lost so much of its dark halo that it would be observed today as a completely old globular cluster with no detectable dark halo. On the other hand, if M54\rq{}s and the surrounding halo was still in the galactic outskirts,  it could have grown into a substantial dwarf galaxy and the central star cluster could well have grown into a much larger nuclear star cluster, possibly even into one containing a detectable black hole. In that sense, M54 can be considered a \lq\lq{}transition\rq\rq{} object.

\section{Discussion and Conclusions}

Our photometric selection using isochrones adjusted to the {\it Gaia} data separates the stars into blue, red (metal rich and younger) and an intermediate, \lq\lq{}green\rq\rq{} population. Each population is kinematically and spatially distinct, although the red and green populations are similar.  A Jeans equation mass analysis using the three populations finds that the mass inside 95 pc is approximately 3.5 times the stellar mass of the M54 star cluster. Beyond 100 pc the mass continues to grow  approximately linearly with distance. The overall rise of mass from 30 to 1000 pc is nearly linear, $M\propto r^{1.1}$, implying an approximately $r^{-2}$ density profile, with a core radius no larger than 30-50 pc.   N-body simulations that start 2.29 Gyr in the past with a star cluster in a Hernquist sphere of masses  $1-3\times 10^8 M_\sun$ and a scale radius of 400 pc provide a good match to most of the properties of the system. The system may have been much more massive when it first encountered the Milky Way, before it spiraled into the current orbit.

Most of the Milky Way halo globular clusters have no significant dark matter detected in their outskirts. However,  $\sim$10\% of the accessible halo clusters have been tentatively identified as being consistent with having a local dark matter halo  \citep{CarlGrill21}. Simulations show that approximately 25-35\% of clusters started in the centers of local halos retain significant dark matter as the Galactic halo assembles \citep{CarlbergK21}.  The M54-Sagittarius system appears to be an intermediary system that would have become a standard globular cluster had it fallen in earlier and its orbit taken it into regions of somewhat stronger tides. On the other hand, if the system had stayed in the outskirts the surrounding dwarf galaxy would likely have continued to form stars to become a more prominent dwarf galaxy today.

The Sagittarius galaxy and its stream contains 8 other high confidence globular clusters \citep{DaCostaArmandroff95,SgrClusters20}  including the distant, high mass NGC 2419 cluster \citep{Ibata13}. Four of the eight are old, metal poor clusters, the other 4 are more metal rich and younger, along the lines of the disk and halo clusters of the Milky Way \citep{SearleZinn78}.  An additional 10-20 lower luminosity clusters are proposed members of Sagittarius \citep{Minniti21}.  If the Milky Way is assembled out of systems similar to Sagittarius, then its  one cluster (M54) in a dark halo of the nine luminous clusters somewhat weakly bolsters the idea that roughly 10\% of disk galaxy clusters retain pre-galactic dark matter halos. If the M54/Sagittarius system is a late-time accretion event that is representative of earlier ones \citep{Massari19,Malhan22}, then the globular clusters in those accretion events may still have one or two clusters with local dark matter halos orbiting freely. 

\begin{acknowledgements}
This research was supported by  NSERC of Canada. 
\end{acknowledgements}

\bibliography{M54}{}
\bibliographystyle{aasjournal}

\end{document}